\documentclass[twocolumn,pra,amsmath,amssymb,superscriptaddress]{revtex4-1}

\usepackage{graphicx}
\usepackage{amssymb}
\usepackage{bm}

\usepackage{color}

\renewcommand{\vec}[1]{\boldsymbol{#1}}
\newcommand{\beq}{\begin{equation}}
\newcommand{\eeq}{\end{equation}}
\newcommand{\bea}{\begin{align}}
\newcommand{\eea}{\end{align}}
\newcommand{\beqa}{\begin{eqnarray}}
\newcommand{\eeqa}{\end{eqnarray}}
\newcommand{\w}{\omega}
\newcommand{\ket}[1]{\left| #1 \right\rangle}
\newcommand{\bra}[1]{\left\langle #1 \right|}

\newcommand{\av}[1]{\langle #1\rangle}
\newcommand{\braket}[2]{\langle #1 | #2\rangle}

\newcommand{\ketbra}[2]{\left|#1\right\rangle\hskip-0.5mm\left\langle #2\right|}

\newcommand{\kv}{ {\bf k} }
\newcommand{\sfrac}[2]{{\textstyle{\frac{ #1}{#2}}}}
\newcommand{\pder}[2]{\frac{\partial #1}{\partial #2} }
\newcommand{\e}{ \mathrm{e} }
\newcommand{\Sys}{ \mathrm{S} }
\newcommand{\Env}{ \mathrm{E} }
\newcommand{\Var}{ \mathrm{V} }
\newcommand{\Pol}{ \mathrm{P} }
\newcommand{\pol}{\mathrm{p}}
\newcommand{\II}{ \mathrm{I} }

\newcommand{\weak}{ \mathrm{w} }

\begin{document}

\title{Modelling exciton-phonon interactions in optically driven quantum dots}
\author{Ahsan Nazir}
\email{ahsan.nazir@manchester.ac.uk}
\affiliation{{Photon Science Institute \& School of Physics and Astronomy, The University of Manchester, Oxford Road, Manchester M13 9PL, United Kingdom}}
\author{Dara P. S. McCutcheon}
\email{daramc@fotonik.dtu.dk}
\affiliation{Department of Photonics Engineering, Technical University of Denmark, Building 345, 2800 Kgs.~Lyngby, Denmark}
\date{\today}

\begin{abstract}
We provide a self-contained review of master equation  
approaches to modelling phonon effects in 
optically driven self-assembled quantum dots. 
Coupling of the (quasi) two-level excitonic system to phonons leads to dissipation and dephasing, 
the rates of which depend on the excitation conditions, intrinsic properties of the QD 
sample, and its  
temperature. We describe several techniques, which include weak-coupling master equations that are perturbative in the exciton-phonon coupling, as well as those based on the polaron transformation that can remain valid for strong phonon interactions. We additionally consider the role of phonons in altering the optical emission characteristics of quantum dot devices, outlining how we must modify standard quantum optics treatments to account for the presence of the solid-state environment. 
\end{abstract}
\maketitle

\section{Introduction}
\label{Introduction}

Self-assembled semiconductor quantum dots (QDs) are small regions 
of a semiconductor alloy hosted in a solid-state matrix composed of a different alloy. 
They are formed, for example, when a thin layer of InAs is deposited on a GaAs substrate. A difference in lattice constants creates strain, 
and above a certain critical thickness the InAs begins to spontaneously nucleate creating 
dome-shaped `islands'~\cite{Notzel1996,jacak98,masumoto02}. The difference in lattice constants also results in a band-gap difference in 
the two materials, and as a consequence charge carriers present in the dot regions 
experience 
confinement in all three spatial dimensions. 
QDs 
are therefore effectively zero-dimensional systems (hence, quantum \emph{dot}), and their charge carriers have discrete 
energy level structures in much the same way as electrons in atoms~\cite{banin99,bayer00,Fujisawa2002}. Similarly, 
they posses transition 
dipole moments, allowing 
for optically mediated excitation, as well as radiative decay and the emission of photons.

These characteristics have led to semiconductor QDs being described as solid-state or artificial atoms. 
Indeed, many of the phenomena more 
traditionally associated to atomic quantum optics have now also been demonstrated with QDs,  
including single and entangled photon emission~\cite{Santori2001,stevenson06,Versteegh2014,Muller2014,Nowak2014,Wei2014_2}, 
two-photon interference~\cite{santori02,Patel2008,flagg10,patel10,Unsleber2015,Gazzano2013,He2013,PhysRevLett.114.067401}, 
driven Rabi oscillations~\cite{kamada01,stievater01,htoon02,borri02,zrenner02,flagg09,ramsay10,ramsay10_2}, and 
resonance fluorescence~\cite{xu07,muller07_long,ates09,flagg09,vamivakas09,ulrich11_short,ulhaq12,konthasinghe12,matthiesen12,Ulhaq2013,Matthiesen2013,Matthiesen2014,Wei2014}. 
Taken in conjunction with their 
solid-state nature and addressability, 
QDs have thus received considerable attention 
for advanced 
technological applications. 
Examples include 
high-quality single photon sources for metrology and secure 
communication~\cite{nguyen11,matthiesen12,konthasinghe12,kiraz04,benjamin09}, as well as 
solid-state and few-emitter lasers~\cite{quilter15}. 
Additionally, due to their relatively long coherence times, confined electron or 
hole spins represent promising realisations of quantum bits for quantum information 
processing~\cite{imamoglu99,piermarocchi02,pazy03,nazir04,atature06,mikkelsen07,gerardot07,berezovsky08,Ramsay2008,press10}, 
and also have the potential to 
act as spin-photon interfaces~\cite{Hu2008,Young2011,Arnold2015}, 
for example within entanglement generation schemes~\cite{Hu2008_2,Lindner2009,Young2013,Gao2012,McCutcheon2014}. 

Besides the many 
prospective applications, a single QD also constitutes an 
archetypal open quantum system~\cite{b+p}, consisting of a small number of controllable 
degrees of 
freedom (the electronic states) coupled to a large uncontrollable environment (the surrounding solid-state lattice). As such, 
QDs offer the further possibility to explore fundamental questions relating to decoherence and thermalisation in 
quantum systems.

Despite evident similarities, the coupling of a QD to its solid-state environment leads to behaviour distinct from that of 
an atom coupled only to the electromagnetic field. Changes in the QD charge configuration perturb the equilibrium positions of the semiconductor lattice ions, 
such that excitons (electron--hole pairs) become sensitive to bulk phonon modes 
and thus also the temperature of the sample. Exciton--phonon coupling has been demonstrated in a number of experimental settings, including 
the observation of emission line broadening in photo-luminescene spectra~\cite{Besombes2001,Peter2004,Favero2003}, 
phonon-induced damping of coherently pumped excitonic Rabi oscillations~\cite{ramsay10,ramsay10_2}, 
sideband linewidth broadening in resonance fluorescence spectra~\cite{ahn05,flagg09,ulrich11_short,Wei2014}, and 
temperature-dependent Rabi frequency renormalisation~\cite{Wei2014,ramsay10_2}.

In this article, we shall review master equation approaches to modelling the effects of exciton-phonon coupling on the dynamical properties of optically driven QDs. Our review is intended to be pedagogical and to that end we have attempted to make it relatively self-contained. We begin in Section~\ref{TheoreticalBackground} by presenting relevant theoretical background, before examining weak exciton-phonon coupling approximations~\cite{machnikowski04,nazir08,rozbicki08,eastham13} in Section~\ref{WeakCoupling}. We also look here at the phenomenological pure-dephasing approximation, highlighting the limited extent to which it can capture the full dynamics predicted by more rigorous methods. Motivated by the breakdown of the weak-coupling treatment in certain regimes, we introduce the polaron transformation~\cite{mahan} in Section~\ref{PolaronTheory} and derive an associated master equation~\cite{mccutcheon10_2}. Though the polaron master equation works well at arbitrary coupling strength for weak-enough optical addressing, it is less successful once the external driving becomes strong due to the unsuitability of the adiabatic (polaron) basis from which it is derived. Section~\ref{VariationalTheory} thus presents a variational extension to the polaron theory which allows both significant driving and strong exciton-phonon interactions to be examined within a consistent framework~\cite{mccutcheon11_2}, and in fact incorporates both the weak-coupling and polaron master equations as limiting cases. We explore the effects of phonons on the optical emission properties of QD devices in Section~\ref{Emission}, with a focus on resonance fluorescence spectra. We finish in Section~\ref{conclusions} with a brief summary and present some technical details in the Appendix. 

It is worth noting 
that master equations constitute only a subset of 
the techniques that have been 
successfully applied to study QD exciton-phonon interactions and related systems~\cite{vagov02,Forstner2003,krugel06,vagov2007,glassl2011,vagov11,Kaer2013,Kaer2013_2}. 
In order to keep the review focussed and concise, however, we shall concentrate only on master equation methods in the following. For the same reasons, as well as simplicity, we consider almost exclusively QDs in the absence of cavity interactions. The subject of phonon effects in cavity-QD systems is a rich and interesting one, with both master equation approaches and various others having been developed in such settings~\cite{wilsonrae02,PhysRevLett.107.193601,roy11,roy12,Kaer2013,Kaer2013_2,vagov2014,ilessmith14arxiv,roychoudhury15,muller15}. Nevertheless, the examples given herein 
are sufficient to elucidate the underlying physics in which we are interested, as well as to illustrate the basic theoretical techniques 
and their many applications. Likewise, polaron physics is widely studied and rapidly developing in numerous different settings~\cite{mitra87,devreese96,alexandrovbook,eminbook}. Again, our main focus in this context will be on the specific application of polaron methods to the dynamics of QD systems.

\section{Theoretical Background}
\label{TheoreticalBackground}

In this section we shall present the theoretical background necessary to describe the dynamics of a coherently driven QD coupled to its surrounding environment. We begin by considering the Hamiltonian for a QD driven by a classical laser field, before outlining how to incorporate quantised phonon and photon interactions. Finally, we introduce the theory of open quantum systems and derive a general master equation with which we shall explore the detailed QD dynamics in subsequent sections.

\subsection{Driven QD Hamiltonian}\label{Hdriven}

We consider a minimal two-level model for the electronic degrees of freedom of a semiconductor QD, 
with basis states defined as $|0\rangle$ and $|X\rangle$. Here, $|0\rangle$ corresponds to the semiconductor vacuum, i.e.~no electrons excited into the conduction band, while $|X\rangle$ represents the exciton state formed from a single electron-hole pair. Such a simplification is widely used 
and, broadly speaking, limits our considerations to situations in which any external driving field is relatively close to resonance with the $\ket{0}$ to $\ket{X}$ transition frequency, such that at the driving strengths considered 
higher lying states are not appreciably excited during the dot dynamics. Likewise, the two-level approximation also limits the ambient temperature such that thermal excitation of higher lying states is strongly suppressed. It is worth bearing in mind that even in such situations, fine structure splitting of the single exciton state can cause further complications~\cite{stace03,hohenester07b}. 
We note, however, that these 
can typically be avoided by carefully aligning the polarisation of the excitation laser in order to predominately excite 
a single exciton transition (see e.g. Refs.~\cite{Wei2014,matthiesen12}). We shall not, therefore, consider extensions beyond the two-level QD model in this review, other than to note that the master equation techniques we develop may be applied more generally to larger systems as well~\cite{stace03,hohenester07b,Kaer2013_2,roy11,roy12,ilessmith14arxiv,kolli11,jang11,pollock13}.

Coherent 
Rabi oscillations may be driven between states $|0\rangle$ and $|X\rangle$ through the action of a classical external laser field, providing a means for coherent control of the QD state. Within the dipole approximation, we consider the Hamiltonian (we set $\hbar=1$ throughout)
\begin{align}\label{dipoledriving}
H_{\rm QD-laser}=\epsilon_X|X\rangle\langle X|-{\mathbf d}\cdot{\mathbf E}(t),
\end{align} 
where $\epsilon_X$ is the energy difference between the QD ground and excited state, ${\mathbf d}$ is the dot dipole operator, and ${\mathbf E}(t)={\bm\epsilon}E_0\cos(\omega_l t)$ is a monochromatic field of frequency $\omega_l$, amplitude $E_0$, and unit polarisation vector ${\bm \epsilon}$. Decomposing the dipole operator in terms of our QD basis states, we may write  ${\mathbf d}=\langle0|{\mathbf d}|X\rangle|0\rangle\langle X|+\langle X|{\mathbf d}|0\rangle|X\rangle\langle 0|$, where we have assumed zero permanent dipole moment and 
used the fact that the dipole operator has odd parity to set $\langle0|{\mathbf d}|0\rangle=\langle X|{\mathbf d}|X\rangle=0$~\cite{koklovett,carmichael,stecknotes}. The field thus drives transitions between the two QD basis states. Assuming the dot dipole matrix element $\langle0|{\mathbf d}|X\rangle$ to be real, we may then write Eq.~(\ref{dipoledriving}) in the form 
\begin{align}\label{drivenQD}
H_{\rm QD-laser}=\epsilon_X|X\rangle\langle X|+\Omega\cos(\omega_lt)(|0\rangle\langle X|+|X\rangle\langle 0|),
\end{align}
where we have defined the Rabi frequency as $\Omega=-E_0{\bm \epsilon}\cdot\langle0|{\mathbf d}|X\rangle$, which quantifies the dot-field coupling strength. As defined here, the Rabi frequency is time-independent, which is appropriate for continuous-wave driving. Of course, it is also possible to consider pulsed excitation by defining a time-varying field amplitude, in which case the Rabi frequency becomes time-dependent as well. 

Even in the continuous-wave case, the oscillatory time dependence in Eq.~(\ref{drivenQD}) is not as straightforward to work with as we would like. However, as stated earlier, we are considering situations in which the driving field is close to resonance. This allows us to simplify the form of Eq.~(\ref{drivenQD}) by making a rotating-wave approximation to remove fast oscillating terms. Transforming to a frame rotating at frequency $\omega_l$ via 
\begin{align}\label{rotatingtrans}
H'=U(t)H_{\rm QD-laser}U^{\dagger}(t)-iU(t)\left(\frac{\partial}{\partial t}U^{\dagger}(t)\right), 
\end{align}
with $U(t)=e^{i\omega_l\ket{X}\bra{X}t}$, gives 
\begin{align}\label{rotatingH}
H'&=(\epsilon_X-\omega_l)\ket{X}\bra{X}\nonumber\\
&+\frac{\Omega}{2}(e^{i\omega_lt}+e^{-i\omega_lt})(e^{-i\omega_lt}\ket{0}\bra{X}+e^{i\omega_lt}\ket{X}\bra{0}).
\end{align}
Neglecting fast oscillating terms at frequencies $\pm2\omega_l$ (the rotating-wave approximation), we then find the simple time-independent form 
\begin{align}\label{hdrivenrwa}
H_{\rm RWA}=\delta\ket{X}\bra{X}+\frac{\Omega}{2}(\ket{0}\bra{X}+\ket{X}\bra{0}),
\end{align}
where we have defined $\delta=\epsilon_X-\omega_l$ as the detuning of the driving frequency from resonance. The validity of the rotating-wave approximation relies upon the inequalities $\delta\ll\omega_l$ and $\Omega\ll\omega_l$ being satisfied. As typical semiconductor band gaps are of the order of $1$~eV (i.e.~$\epsilon_X\sim1$~eV), we can safely explore detunings and Rabi frequencies up to meV scales without the rotating-wave approximation breaking down.

\subsection{Exciton-phonon interactions}

Having outlined how we may describe the optical control of a QD in isolation, we would now like to consider 
interactions with the large number of 
surrounding environmental degrees of freedom that will inevitably be present in any real QD sample. The aim of this section will be to justify an appropriate form for the Hamiltonian governing QD exciton-phonon interactions (rather than a rigorous derivation), the dynamical treatment of which forms the main focus for much of the remaining review. Coupling between the QD and quantised electromagnetic field modes will be considered subsequently.

Lattice vibrations, or phonons, are ubiquitous in solid-state systems. Within the harmonic approximation the free phonon Hamiltonian takes the form
\begin{align}\label{freephohon}
H_{\rm ph}=\sum_{\bf k}\omega_{\bf k}b_{\bf k}^{\dagger}b_{\bf k},
\end{align}
where $b_{\bf k}^{\dagger}$ ($b_{\bf k}$) are creation (annihilation) operators for modes of wavevector ${\bf k}$ and frequency $\omega_{\bf k}$. This Hamiltonian is obtained by 
considering small displacements from equilibrium in a three-dimensional array of ions with nearest neighbour interactions, truncating the interactions at second order in the displacements, then defining and quantising normal modes~\cite{koklovett,b+f,nitzanbook}. 

The excitation of an electron from the valence to the conduction band modifies the charge configuration within the semiconductor crystal, hence altering the equilibrium positions of the lattice ions and giving rise to a coupling between excitons and lattice phonons. Assuming the interaction between an electron and an ion at positions ${\bf r}$ and ${\bf R}$, respectively, depends only on their separation, we may write 
\begin{align}\label{heion}
H_{\rm e-ion}=\int \mathrm{d}{\bf r}\hat{\varrho}({\bf r})\sum_{m}V_{\rm e-ion}({\bf r}-{\bf R}_m),
\end{align} 
where we integrate over the electron charge density operator in the solid 
$\hat{\varrho}({\bf r})=\sum_{jj'}\psi_j^*({\bf r})\psi_{j'}({\bf r})c_j^{\dagger}c_{j'}$ and the sum runs over all 
ions. Here, an electron with wavefunction $\psi_j({\bf r})$ has creation (annihilation) operator $c^{\dagger}_j$ ($c_j$).  
In order to make the ion displacements explicit, we now decompose their positions as 
${\bf R}_m={\bf R}_m^0+{\bf Q}_m$, where ${\bf R}_m^0$ is the ion equilibrium position 
and ${\bf Q}_m$ is the ion displacement (assumed small). Applying a Taylor expansion then yields 
\begin{align}\label{hionexpand}
\sum_m V_{\rm e-ion}\approx\sum_{m}\left[V({\bf r}-{\bf R}_m^0)-{\bf Q}_m\cdot\vec{\nabla} V({\bf r}-{\bf R}_m^0)\right],
\end{align} 
where we have ignored terms of order ${\bf Q}_m^2$ and higher. The first term in Eq.~(\ref{hionexpand}) is simply the periodic potential experienced by the electrons in the unperturbed lattice, which gives rise to Bloch functions. 
As we have already defined our excitonic basis, we do not need to consider this term further. The second term, however, constitutes a linear electron-phonon interaction brought about by the ion displacements:
\begin{align}\label{hep}
V_{\rm ep}=-\sum_{m}{\bf Q}_m\cdot \vec{\nabla} V({\bf r}-{\bf R}_m^0).
\end{align} 
The free phonon Hamiltonian is defined in terms of a sum over wavevectors, hence we consider the Fourier decomposition $V({\bf r})=N^{-1}\sum_{\bf q}v({\bf q})e^{i{\bf q}\cdot{\bf r}}$ and write
\begin{align}\label{hepfourier}
V_{\rm ep}&=-\frac{i}{N}\sum_{{\bf q}}\sum_{m}{\bf Q}_me^{-i{\bf q}\cdot{\bf R}_m^0}\cdot {\bf q}v({\bf q})e^{i{\bf q}\cdot{\bf r}}\nonumber\\
&=-\frac{i}{\sqrt{N}}\sum_{{\bf q}}{\bf Q}_{\bf q}\cdot {\bf q}v({\bf q})e^{i{\bf q}\cdot{\bf r}},
\end{align} 
where ${\bf Q}_{\bf q}=(1/\sqrt{N})\sum_{m}{\bf Q}_me^{-i{\bf q}\cdot{\bf R}_m^0}$. 
The phonon wavevectors ${\bf k}$ are defined only within the first Brillouin zone~\cite{koklovett,b+f}, whereas the vectors ${\bf q}$ are defined everywhere. We can relate the two through the reciprocal lattice vector ${\bf G}$. However, as we are considering only low energy excitations of our system, we shall assume that they couple only to states within the first Brillouin zone, and thus simply replace ${\bf q}$ by ${\bf k}$. The displacement operator ${\bf Q_k}$ may then be written in terms of phonon creation and annihilation operators as ${\bf Q_k}=(i/\sqrt{2m\omega_{{\bf k}}})(b_{{\bf k}}+b^{\dagger}_{-{\bf k}})$ for ions of mass $m$, yielding
\begin{align}\label{hepquant}
V_{\rm ep}&=\sum_{{\bf k}}\frac{1}{\sqrt{2mN\omega_{{\bf k}}}}(b_{\bf k}+b^{\dagger}_{-\bf k})|{\bf k}|v({\bf k})e^{i{\bf k}\cdot{\bf r}},
\end{align} 
where we consider an isotropic system such that $V_{\mathrm{ep}}$ is only non-zero for longitudinally polarised phonons (i.e.~polarisation parallel to ${\bf k}$). From Eq.~(\ref{heion}) we then find that the electron-phonon interaction Hamiltonian can be written as
\begin{align}\label{Helecphonon}
H_{\rm ep}=\sum_{{\bf k}}M_{\bf k}\hat{\varrho}({\bf k})(b_{\bf k}+b^{\dagger}_{-\bf k}),
\end{align}
with $\hat{\varrho}({\bf k})=\int \mathrm{d}{\bf r}\hat{\varrho}({\bf r})e^{i{\bf k}\cdot{\bf r}}$ and $M_{\bf k}=|{\bf k}|v({\bf k})/\sqrt{2mN\omega_{{\bf k}}}$. For the arsenide systems in which we are interested the dominant coupling mechanism is known as the deformation potential, as long as the electron and hole wavefunction overlap is large such that piezoelectric effects are strongly suppressed~\cite{krummheuer02,calarco03,erik}.
To lowest order the deformation potential coupling can be treated by the simple phenomenological approximation of replacing $v({\bf k})$ by experimentally determined constants $D_c$ and $D_v$ relating to the conduction and valence bands, respectively~\cite{mahan,koklovett}.

In the two-level QD approximation we consider only the states $|0\rangle$ and $|X\rangle$, with wavefunctions $\psi_0({\bf r})$ and $\psi_{X}({\bf r})$, respectively. We define the form factors $\varrho_{00}({\bf k})=\langle0|\hat{\varrho}({\bf k})|0\rangle=\int \mathrm{d}{\bf r}|\psi_0({\bf r})|^2e^{i{\bf k}\cdot{\bf r}}$ (electron in the valence band) and $\varrho_{{XX}}({\bf k})=\langle X|\hat{\varrho}({\bf k})|X\rangle=\int \mathrm{d}{\bf r}|\psi_X({\bf r})|^2e^{i{\bf k}\cdot{\bf r}}$ (electron in the conduction band). The ground to excited state splitting $\epsilon_X$ is much larger than the phonon frequencies under consideration (up to meV order), hence the off-diagonal terms are expected to be small and we shall neglect them~\footnote{Within the two-level QD approximation, provided that the frequency dependence of the diagonal and off-diagonal terms is the same, we could account for off-diagonal couplings by rotating the interaction term back into a diagonal form. In this rotated basis, we could then employ the methods outlined for the diagonal coupling case herein. For the effect of off-diagonal couplings to states outside the two-level basis, see Ref.~\cite{muljarov04}.}. 
Subtracting a term proportional to the identity, we then obtain  
\begin{align}\label{Hexcitonphonon}
H_{\rm ep}=|X\rangle\langle X|\sum_{{\bf k}}g_{\bf k}(b_{\bf k}+b^{\dagger}_{-\bf k}),
\end{align}
where
\begin{align}\label{gk}
g_{\bf k}=M_{\bf k}^{(X)}\varrho_{{XX}}({\bf k})-M_{\bf k}^{(0)}\varrho_{00}({\bf k}),
\end{align}
defines the coupling constants, with $M_{\bf k}^{(0)}=|{\bf k}|D_v/\sqrt{2mN\omega_{{\bf k}}}$ and $M_{\bf k}^{(X)}=|{\bf k}|D_c/\sqrt{2mN\omega_{{\bf k}}}$. A simple phenomenological model for the QD wavefunctions can be given within the envelope function approximation by assuming spherically symmetric parabolic potentials for both the conduction and valence bands. This leads to $\psi_{j}({\bf r})=(d_j\sqrt{\pi})^{-3/2}\exp{(-r^2/2d_j^2)}$, for $j=\{0,X\}$, where $d_j$ characterises the size of the wavefunction. Hence $\varrho_{{jj}}({\bf k})=\exp{(-d_j^2|{\bf k}|^2/4)}$, which allows us to write the exciton-phonon interaction Hamiltonian in the usual form 
\begin{align}\label{Hexcitonphonon}
H_{\rm ep}=|X\rangle\langle X|\sum_{{\bf k}}g_{\bf k}(b_{\bf k}+b^{\dagger}_{\bf k}),
\end{align}
where $g_{\bf k}=|{\bf k}|/\sqrt{2mN\omega_{{\bf k}}}[D_ce^{-d_X^2|{\bf k}|^2/4}-D_ve^{-d_0^2|{\bf k}|^2/4}]$.

In what follows it will be useful to introduce the exciton-phonon spectral density, defined as
\begin{align}\label{phononSDsum}
J_{\rm ph}(\omega)=\sum_{\bf k}g_{\bf k}^2\delta(\omega-\omega_{\bf k})=D_{\rm ph} (\omega) g^2(\omega),
\end{align}
which is a measure of the exciton-phonon coupling strength $g(\omega)$ in the continuum limit, weighted by the phonon density of states $D_{\rm ph} (\omega)$.
Taking $d_X=d_0=d$ for simplicity, converting the sum into an integral via $\sum_{\bf k}\rightarrow V/(2\pi)^3\int \mathrm{d}{\bf k}$, 
and assuming linear dispersion $\omega_{\bf k}=c|\bf k|$ with $c$ the speed of sound, we find 
\begin{align}\label{phononSD}
J_{\rm ph}(\omega)=\alpha\omega^3e^{-\omega^2/\omega_c^2},
\end{align}
where $\alpha=V(D_c-D_v)^2/(4\pi^2mNc^5)$ and the cut-off frequency is $\omega_c=\sqrt{2}c/d$. We shall consider the phonon spectral density to take this form throughout. 


\subsection{Exciton-photon interactions} 

In addition to interactions with lattice phonons, QDs are also coupled to surrounding electromagnetic field modes. In the case of the continuum (free) electromagnetic field this gives rise to spontaneous photon emission processes via electron-hole recombination, limiting the QD exciton lifetime. In contrast, for QDs within electromagnetic cavities, interactions between the dot and the discrete field modes can be coherent, resulting in joint exciton-photon eigenstates known as polaritons. Loss of excitation then occurs via non-cavity modes,  cavity leakage, or both. 

In Section~\ref{Hdriven} we considered the coupling between a two-level QD and a classical field within the dipole approximation [see Eq.~(\ref{dipoledriving})]. 
We shall now extend this treatment to the case of a QD coupled to a quantised electromagnetic field. 
Let us begin by considering a QD-cavity system in which the geometry is such that a single field mode dominates. 
Within the dipole approximation, we again write our QD-field interaction as
\begin{align}\label{Hdipole2}
H_{\rm dipole}=-{\bf d}\cdot{\bf E},
\end{align}
where now for a uniform cavity 
\begin{align}\label{Equantum}
{\bf E}={\bm \epsilon}\sqrt{\frac{\omega_0}{2\epsilon_0V}}(a+a^{\dagger})
\end{align}
is written in terms of quantised field mode operators $a$ and $a^{\dagger}$, which define the harmonic cavity Hamiltonian
\begin{align}\label{Hcavity}
H_{\rm cav}=\omega_0 a^{\dagger}a.
\end{align}
Here, $\omega_0$ is the cavity frequency, $V$ is the quantisation (cavity) volume, and ${\bm\epsilon}$ is the mode polarisation vector at the QD location. 
In analogy to the Rabi frequency in Eq.~(\ref{drivenQD}), we define the coupling constant in the present case as
\begin{align}\label{gcouplingdefn}
u=-\sqrt{\frac{\omega_0}{2\epsilon_0V}}{\bm{\epsilon}}\cdot\langle0|{\bf d}|X\rangle,
\end{align}
which allows us to write
\begin{align}\label{Hdipole3}
H_{\rm dipole}=u(|0\rangle\langle X|+|X\rangle\langle0|)(a+a^{\dagger}).
\end{align}
Notice that the coupling scales as $u\sim1/\sqrt{V}$, meaning that the smaller the cavity and more well confined the mode, the stronger the coupling to the QD. 
The full QD-cavity Hamiltonian then becomes 
\begin{align}\label{HRabi}
H_{\rm QD-cav}&=\epsilon_X|X\rangle\langle X|+\omega_0a^{\dagger}a\nonumber\\
&+u(|0\rangle\langle X|+|X\rangle\langle0|)(a+a^{\dagger}),
\end{align}
which is also known as the Rabi Hamiltonian.

As in the case of a classical field, for a QD and cavity close to resonance we may perform a rotating-wave-approximation, 
provided that the dot-cavity coupling is not too strong. These assumptions are usually well satisfied in actual QD-cavity systems, 
and permit us to ignore the terms $|0\rangle\langle X|a$ and $|X\rangle\langle0|a^{\dagger}$ which are then far off-resonant. 
This leads to a QD-cavity Hamiltonian in Jaynes-Cummings form 
\begin{align}\label{HJC}
H_{\rm QD-cav}=\epsilon_X|X\rangle\langle X|+\omega_0a^{\dagger}a+u(|0\rangle\langle X|a^{\dagger}+|X\rangle\langle0|a).
\end{align}
As the Jaynes-Cummings Hamiltonian preserves excitation number ($a^{\dagger}a+\ketbra{X}{X}$) it may be diagonalised 
straightforwardly to give manifolds of entangled light-matter (dressed) states. For example, on 
resonance ($\epsilon_X=\omega_0$) the QD-cavity eigenstates can be expressed as
\begin{align}\label{JCresonance}
\frac{1}{\sqrt{2}}|0,n\rangle\pm|X,n-1\rangle,
\end{align}
split by the Rabi frequency $2u\sqrt{n}$, where $|n\rangle$ denotes a cavity Fock (number) state 
satisfying $a^{\dagger}a|n\rangle=n|n\rangle$. Note that as the Rabi frequency grows with $\sqrt{n}$, in the classical limit $n\gg1$ adjacent manifolds are split by almost the same amount. For small photon numbers, however, the splittings are strongly $n$-dependent.

To generalise to the case of a multimode field,
\begin{align}\label{multimodefree}
H_{\rm field}=\sum_{\bf q}\nu_{\bf q}a_{\bf q}^{\dagger}a_{\bf q},
\end{align}
 we may write our QD-photon interaction Hamiltonian (within the rotating-wave approximation) as
\begin{align}\label{multimodeint}
H_{\rm int}=\sum_{\bf q}u_{\bf q}(|0\rangle\langle X|a_{\bf q}^{\dagger}+|X\rangle\langle0|a_{\bf q}),
\end{align}
where 
\begin{align}\label{gcouplingdefn}
u_{\bf q}=-\sqrt{\frac{\omega_{\bf q}}{2\epsilon_0V}}{\bm{\epsilon}_{\bf q}}\cdot\langle0|{\bf d}|X\rangle,
\end{align}
is the coupling strength for mode ${\bf q}$ and the sum over mode polarisations is implicit. As stated previously, the QD-mode coupling increases for well confined modes. Thus, in the vacuum field case relevant to spontaneous emission processes, the coupling to each individual mode is expected to be small even if their combined effect is significant. Hence, we do not need to consider dressed states in this situation, and can instead treat the QD-photon coupling perturbatively. This can be achieved in the context of open quantum systems theory by deriving a master equation governing the QD evolution under the influence of the photon environment. As we shall be using master equations extensively throughout the rest of this review to capture the effects of both photon and phonon environments, in the next section we shall briefly overview their derivation and the basics of open quantum systems. 

\subsection{Open quantum systems and master equations}

We define an open quantum system $\mathrm{S}$ as a subsystem of a larger combined system $\mathrm{S+E}$. 
Here, $\mathrm{E}$ represents another quantum system, the environment, to which $\mathrm{S}$ is coupled; 
for example, an exciton (the system) interacting with the vibrational modes of a solid or with the surrounding electromagnetic 
field (the environment). We shall assume that the combined evolution of the system-plus-environment ($\mathrm{S+E}$) is closed, 
and so follows unitary Hamiltonian dynamics. The state of the system $\mathrm{S}$, however, will evolve not only according to 
its own internal Hamiltonian, but also due to interactions with the environment $\mathrm{E}$. In this situation, it is generally 
not possible to represent the system dynamics by a unitary evolution operator acting on $\mathrm{S}$ alone. 
Instead, we shall derive a master equation governing the dynamics of the \emph{reduced} density operator 
$\rho_{\rm S}(t)={\rm Tr}_{\rm E}[\rho(t)]$, which represents the system state once we trace out the environmental 
degrees of freedom. The reduced density operator describes all accessible information about the system $\mathrm{S}$. 

We begin our master equation derivation by writing the system-environment Hamiltonian as
\begin{equation}\label{sysenvH}
H=H_{\rm S}+H_{\rm I}+H_{\rm E},
\end{equation}
where we assume that the interaction Hamiltonian $H_{\rm I}$ is the only part that involves both system and environment degrees of freedom. It will be this part of the Hamiltonian that is treated as a perturbation. To make this more explicit, we now take $H_0=H_{\rm S}+H_{\rm E}$, and move into the interaction picture
\begin{equation}
\tilde{H}_{\rm I}(t)=e^{i(H_{\rm S}+H_{\rm E})t}H_{\rm I}e^{-i(H_{\rm S}+H_{\rm E})t},
\end{equation}
where we set $t_0=0$. Here, a tilde is used to represent an operator that has been transformed into the interaction picture, i.e. 
$\tilde{O}(t)=e^{iH_0t}Oe^{-iH_0t}$. As the combined system and environment is closed, within the interaction picture the system-environment density operator evolves according to
\begin{equation}\label{lvonnintmaster}
\frac{\mathrm{d}\tilde{\rho}(t)}{\mathrm{d}t}=-i[\tilde{H}_{\rm I}(t),\tilde{\rho}(t)].
\end{equation}
This equation has the formal solution
\begin{equation}\label{lvonnintsol}
\tilde{\rho}(t)=\rho(0)-i\int_0^t\mathrm{d}s[\tilde{H}_{\rm I}(s),\tilde{\rho}(s)],
\end{equation}
which we may substitute back into Eq.~(\ref{lvonnintmaster}) to give
\begin{equation}\label{lvonnintmaster2}
\frac{\mathrm{d}\tilde{\rho}(t)}{\mathrm{d}t}=-i[\tilde{H}_{\rm I}(t),\rho(0)]-\int_0^t\mathrm{d}s[\tilde{H}_{\rm I}(t),[\tilde{H}_{\rm I}(s),\tilde{\rho}(s)]].
\end{equation}
%
Following Ref.~\onlinecite{wallsmilburn} we iterate this solution to generate a series expansion in terms of $\rho(0)$: 
\begin{align}
\tilde{\rho}(t)&=\rho(0)+\sum_{n=1}^{\infty}(-i)^n\int_{0}^{t}\mathrm{d}t_1\int_{0}^{t_1}\mathrm{d}t_2\cdots\int_{0}^{t_{n-1}}\mathrm{d}t_n\nonumber\\
&\times[\tilde{H}_{\mathrm I}(t_1),[\tilde{H}_{\mathrm I}(t_2),\cdots[\tilde{H}_{\mathrm I}(t_n),\rho(0)]]\cdots].
\end{align}
Taking a trace over the environmental degrees of freedom, we find
\begin{align}\label{reducedexpand}
\tilde{\rho}_{\rm S}(t)&=\rho_{\rm S}(0)+\sum_{n=1}^{\infty}(-i)^n\int_{0}^{t}\mathrm{d}t_1\int_{0}^{t_1}\mathrm{d}t_2\cdots\int_{0}^{t_{n-1}}\mathrm{d}t_n\nonumber\\
&\times{\mathrm {Tr}}_{\mathrm E}[\tilde{H}_{\mathrm I}(t_1),[\tilde{H}_{\mathrm I}(t_2),\cdots[\tilde{H}_{\mathrm I}(t_n),\rho_S(0)\rho_E(0)]]\cdots],
\end{align}
where we assume $\rho(0)=\rho_{\mathrm S}(0)\rho_{\mathrm E}(0)$ factorises initially. We may write Eq.~(\ref{reducedexpand}) in the form
\begin{align}\label{reducedexpand2}
\tilde{\rho}_{\rm S}(t)&=(1+W_1(t)+W_2(t)+\cdots)\rho_{\rm S}(0),\nonumber\\
&=W(t)\rho_{\rm S}(0),
\end{align}
where 
\begin{align}
W_n(t)&=(-i)^n\int_{0}^{t}\mathrm{d}t_1\int_{0}^{t_1}\mathrm{d}t_2\cdots\int_{0}^{t_{n-1}}\mathrm{d}t_n\nonumber\\
&\times{\rm Tr}_{\rm E}[\tilde{H}_{\rm I}(t_1),[\tilde{H}_{\rm I}(t_2),\cdots[\tilde{H}_{\rm I}(t_n),(\cdot)\rho_E(0)]]\cdots],
\end{align}
are superoperators acting on the initial system density operator. 
Differentiating with respect to time, we have
\begin{align}\label{masterexpand}
\frac{\mathrm{d} \tilde{\rho}_{\rm S}(t)}{\mathrm{d}t}&=(\dot{W}_1(t)+\dot{W}_2(t)+\cdots)\rho_{\rm S}(0),\nonumber\\
&=(\dot{W}_1(t)+\dot{W}_2(t)+\cdots)W(t)^{-1}\tilde{\rho}_{\rm S}(t),
\end{align} 
where we have used Eq.~(\ref{reducedexpand2}) and assumed that $W(t)$ is invertible. 
Usually, it is convenient (and possible) to define the interaction Hamiltonian such that ${\rm Tr}_{\rm E}[\tilde{H}_{\rm I}(t)\rho_{\rm E}(0)]=0$, which means that $W_1(t)=0$ as well. Thus, to second order, Eq.~(\ref{masterexpand}) becomes
\begin{align}\label{MEinteraction}
\frac{\mathrm{d}\tilde{\rho}_{\rm S}(t)}{\mathrm{d}t}&=\dot{W}_2(t)\tilde{\rho}_{\rm S}(t),\nonumber\\
&=-\int_0^{t}\mathrm{d}t_1{\rm Tr}_{\rm E}[\tilde{H}_{\rm I}(t),[\tilde{H}_{\rm I}(t_1),\tilde{\rho}_{\rm S}(t)\rho_{\rm E}(0)]],
\end{align} 
which when replacing $t_1\rightarrow t-\tau$ and moving back into the Schr\"odinger picture gives 
\begin{align}\label{SchrodingerME}
\frac{\mathrm{d} {\rho}_{\rm S}(t)}{\mathrm{d}t}&=-i[H_{\rm S},\rho_{\rm S}(t)]\nonumber\\
&-\int_0^{t}\mathrm{d}\tau{\rm Tr}_{\rm E}[{H}_{\rm I},[\tilde{H}_{\rm I}(-\tau),{\rho}_{\rm S}(t)\rho_{\rm E}(0)]].
\end{align}
We now decompose our interaction Hamiltonian as
\begin{align}\label{intdecomp}
\tilde{H}_{\rm I}(t)=\sum_{i}\tilde{A}_{i}(t)\otimes\tilde{B}_{i}(t),
\end{align}
where $\tilde{A}_{i}(t)=e^{iH_{\rm S}t}A_{i}e^{-iH_{\rm S}t}$ and $\tilde{B}_{i}(t)=e^{iH_{\rm E}t}B_{i}e^{-iH_{\rm E}t}$ are system and environment operators, respectively. 
Substituting into Eq.~(\ref{SchrodingerME}) we find 
\begin{align}\label{SchrodingerME2}
\frac{\mathrm{d} {\rho}_{\rm S}(t)}{\mathrm{d}t}&=-i[H_{\rm S},\rho_{\rm S}(t)]\nonumber\\
&-\sum_{ij}\int_0^{t}\mathrm{d}\tau\Big(C_{ij}(\tau)[A_{i},\tilde{A}_{j}(-\tau){\rho}_{\rm S}(t)]\nonumber\\
&+C_{ji}(-\tau)[\rho_{\rm S}(t)\tilde{A}_{j}(-\tau),A_{i}]\Big).
\end{align}
Here, we have defined the environmental correlation functions
\begin{align}\label{envcorr}
C_{ij}(\tau)={\rm Tr}_{\rm E}[\tilde{B}_{i}(\tau)B_{j}\rho_{\rm E}(0)],
\end{align}
and have also made use of the fact that we shall consider only stationary (equilibrium) initial environmental states which satisfy $[H_{\rm E},\rho_{\rm E}(0)]=0$. 
The time-dependence of the system operators $\tilde{A}_{i}(\tau)$ may be made explicit by considering the Fourier decomposition
\begin{align}\label{fourieroperator}
\tilde{A}_{i}(\tau)=\sum_{\zeta}e^{-i\zeta\tau}A_{i}(\zeta),
\end{align}
where the sum extends over all system eigenvalue differences. 
If the environmental correlation functions are short lived we may extend the upper limit of integration in Eq.~(\ref{SchrodingerME2}) to infinity to give the Markovian master equation 
\begin{align}\label{SchrodingerMEMarkov}
\frac{\mathrm{d} {\rho}_{\rm S}(t)}{\mathrm{d}t}&=-i[H_{\rm S},\rho_{\rm S}(t)]\nonumber\\
&-\sum_{ij}\int_0^{\infty}\mathrm{d}\tau\Big(C_{ij}(\tau)[A_{i},\tilde{A}_{j}(-\tau){\rho}_{\rm S}(t)]\nonumber\\
&+C_{ji}(-\tau)[\rho_{\rm S}(t)\tilde{A}_{j}(-\tau),A_{i}]\Big).
\end{align}
Note that this is equivalent to taking the lower limit in Eq.~(\ref{MEinteraction}) to $-\infty$, such that it no longer contains any reference to a particular preparation at $t=0$. 
The evolution then depends only on the present state of the system, rather than its history, as expected for a Markovian process.

For a decomposition of the interaction Hamiltonian in terms of Hermitian operators, i.e.~$\tilde{A}_i(\tau)=\tilde{A}_i^{\dagger}(\tau)$ and $\tilde{B}_i(\tau)=\tilde{B}_i^{\dagger}(\tau)$, we may use Eq.~({\ref{fourieroperator}}) to write the master equation in a slightly neater form  
\begin{align}
\frac{\mathrm{d}{\rho_{\rm S}}(t)}{\mathrm{d}t}=-&i[H_S,\rho_{\rm S}(t)]\nonumber\\
-&\sfrac{1}{2}\sum_{ij}\sum_\zeta \gamma_{ij}(\zeta)[A_{i},A_{j}(\zeta)\rho_{\rm S}(t)-\rho_{\rm S}(t)A_{j}^{\dagger}(\zeta)]\nonumber\\
-&i\sum_{ij}\sum_\zeta S_{ij}(\zeta)[A_{i},A_{j}(\zeta)\rho_{\rm S}(t)+\rho_{\rm S}(t)A_{j}^{\dagger}(\zeta)],
\label{rhodotD2}
\end{align}
where $A_j^{\dagger}(\zeta)=A_j(-\zeta)$ and we have defined the rates and energy shifts as the real and imaginary components of the response functions
\begin{align}
K_{ij}(\zeta)=&\int_0^{\infty} \mathrm{d}\tau C_{ij}(\tau)\e^{i\zeta\tau}\nonumber\\
=&\frac{1}{2}\gamma_{ij}(\zeta)+i S_{ij}(\zeta),
\label{KDefinition}
\end{align}
such that $\gamma_{ij}(\zeta)=2\mathrm{Re}[K_{ij}(\zeta)]$ and 
$S_{ij}(\zeta)=\mathrm{Im}[K_{ij}(\zeta)]$. 

\subsection{Spontaneous emission}\label{sebackground}

As an example of the utility of the master equation formalism, we return to the case of a QD (weakly) 
coupled to the continuum vacuum photon field outlined previously. Here, we have $H_{\rm S}=\epsilon_X|X\rangle\langle X|$, $H_{\rm E}=\sum_{\bf q}\nu_{\bf q}a_{\bf q}^{\dagger}a_{\bf q}$, and $H_{\rm I}=\sum_{\bf q}u_{\bf q}(\sigma_-a_{\bf q}^{\dagger}+\sigma_+a_{\bf q})$, where we have defined $\sigma_-=|0\rangle\langle X|$ and $\sigma_+=|X\rangle\langle0|$. Moving into the interaction picture, we write
\begin{align}\label{Hintvacuum}
\tilde{H}_{\rm I}(t)=\tilde{A}_1(t)\tilde{B}_{1}(t)+\tilde{A}_2(t)\tilde{B}_{2}(t),
\end{align}
where $\tilde{A}_1(t)=\sigma_-e^{-i\epsilon_Xt}$, $\tilde{A}_2(t)=\sigma_+e^{i\epsilon_Xt}$, $\tilde{B}_1(t)=\sum_{\bf q}u_{\bf q}a_{\bf q}^{\dagger}e^{i\nu_{\bf q}t}$, and $\tilde{B}_2(t)=\sum_{\bf q}u_{\bf q}a_{\bf q}e^{-i\nu_{\bf q}t}$ (note that this is not a decomposition in terms of Hermitian operators). Taking the initial environmental state to be the multimode vacuum, we find that the only non-zero bath correlation function is $C_{21}(\tau)=\sum_{\bf qq'}u_{\bf q}u_{\bf q'}e^{-i\nu_{\bf q}\tau}\langle a_{\bf q}a_{\bf q'}^{\dagger}\rangle$, where the vacuum expectation is $\langle a_{\bf q}a_{\bf q'}^{\dagger}\rangle=\delta_{\bf qq'}$. Thus, $C_{21}(\tau)=\sum_{\bf q}u_{\bf q}^2e^{-i\nu_{\bf q}\tau}$, which in the continuum limit becomes
\begin{align}\label{continuumlimit}
C_{21}(\tau)=\int_0^{\infty}\mathrm{d}\nu J_{\mathrm{pt}}(\nu)e^{-i\nu\tau}.
\end{align}
Here, we have defined the photon spectral density
\begin{align}\label{specdensdefn}
J_{\mathrm{pt}}(\nu)=\sum_{\bf q}u_{\bf q}^2\delta(\nu-\nu_{\bf q})=D_{\mathrm{pt}}(\nu)u(\nu)^2,
\end{align}
which, in analogy with the phonon spectral density, is a measure of the exciton--photon coupling strength weighted by the 
electromagnetic density of states $D_{\mathrm{pt}}(\nu)$. Substituting into Eq.~(\ref{SchrodingerMEMarkov}) then gives
\begin{align}\label{opticalME1}
\frac{\mathrm{d} {\rho}_{\rm S}(t)}{\mathrm{d}t}&=-i[\epsilon_X|X\rangle\langle X|,\rho_{\rm S}(t)]\nonumber\\
&-\int_0^{\infty}\mathrm{d}\tau\int_0^{\infty}\mathrm{d}\nu J_{\mathrm{pt}}(\nu)\Big(e^{i(\epsilon_X-\nu)\tau}[\sigma_+,\sigma_-{\rho}_{\rm S}(t)]\nonumber\\
&+e^{-i(\epsilon_X-\nu)\tau}[\rho_{\rm S}(t)\sigma_+,\sigma_-]\Big).
\end{align}
To evaluate the integrals we use the relation
\begin{align}\label{diracreln}
\int_{0}^{\infty}\mathrm{d}\tau e^{\pm i\nu\tau}=\pi\delta(\nu)\pm iP\frac{1}{\nu},
\end{align}
where $P$ stands for the Principal Value, 
and thus obtain the standard form for the optical master equation~\cite{b+p,carmichael}
\begin{align}\label{opticalME2}
\frac{\mathrm{d} {\rho}_{\rm S}(t)}{\mathrm{d}t}&=-i[\epsilon_X'\sigma_+\sigma_-,\rho_{\rm S}(t)]\nonumber\\
&+\gamma(\epsilon_X)\left(\sigma_-\rho_{\rm S}(t)\sigma_+-(1/2)\{\sigma_+\sigma_-,\rho_{\rm S}(t)\}\right),
\end{align}
with QD spontaneous emission rate 
\begin{align}\label{serate}
\gamma(\epsilon_X)=2\pi J_{\mathrm{pt}}(\epsilon_X),
\end{align} 
and (Lamb) shifted excitation energy $\epsilon_X'=\epsilon_X+P\int_0^{\infty}\mathrm{d}\nu J_{\mathrm{pt}}(\nu)/(\epsilon_X-\nu)$.

\section{QD Dynamics - Weak-Coupling}
\label{WeakCoupling}

With the necessary theoretical background established, we now begin our 
investigation 
of the optically driven QD dynamics within what we term  
weak (phonon) coupling theory~\cite{nazir08,machnikowski04,rozbicki08,eastham13}. This treatment is based on the Markovian master equation technique introduced in Section~\ref{TheoreticalBackground}, 
taking the exciton-phonon interaction term of Eq.~(\ref{Hexcitonphonon}) as a perturbation. We shall see that as long as: (i) the exciton-phonon coupling strength does not become too large;
(ii) the temperature is low, in a sense that we shall clarify below; and (iii) the driving does not induce dynamics so fast that they are unresolved within the Markov approximation, then weak-coupling theory is sufficient to describe the QD dynamics for typically relevant experimental parameters. 
In fact, 
the theory has been successfully applied to infer and interpret the influence of phonons in experimental QD exciton Rabi rotation data, for example see Refs.~\cite{ramsay10,ramsay10_2}.

For clarity, it is worth restating the Hamiltonian for the  
complete QD exciton-phonon system. 
We consider a QD driven by a classical laser field 
with Rabi frequency $\Omega$, coupled to phonons via the deformation potential. 
As such, our system Hamiltonian is given by Eq.~({\ref{hdrivenrwa}}), our environment Hamiltonian by Eq.~({\ref{freephohon}}), and the interaction Hamiltonian by 
Eq.~({\ref{Hexcitonphonon}}). 
The combined Hamiltonian in the rotating frame is thus
\beq
H=H_\Sys+H_\II+H_\Env,
\label{HTotal}
\eeq
where the various contributions are 
%
\begin{align}
H_\Sys & =  \delta\ketbra{X}{X}+\frac{\Omega}{2}(\ketbra{0}{X}+\ketbra{X}{0}),\label{HS}\vphantom{\sum_\kv}\\
H_\II & =  \ketbra{X}{X}\sum_\kv g_\kv (b_\kv^{\dagger}+b_\kv),\label{HI}\\
H_\Env & =  \sum_\kv \w_\kv b_\kv^{\dagger}b_\kv,\label{HE}
\end{align}
with definitions as given in Section~\ref{TheoreticalBackground}.

\subsection{Pure-dephasing approximation}
\label{PureDephasingApproximation}

Before deriving the full weak-coupling 
master equation, it is first instructive to consider a simple 
phenomenological pure-dephasing model for the phonon influence, which is sometimes employed. This model is motivated by the form of 
$H_{\rm I}$ in Eq.~({\ref{HI}}). The interaction 
between the system and environment takes place through the diagonal QD operator $\ketbra{X}{X}$. As such, we may expect 
that for particular classes of Hamiltonians, system operators which commute with this term will have preserved expectation values, while non-commuting 
operators will have decaying expectation values. Physically, this means that QD populations are constants of motion, 
whereas the coherences dephase. We shall see in Section~\ref{PolaronTheory} 
that in the zero driving limit ($\Omega\to 0$) this can rigorously be shown to be the case, though it is not true for finite $\Omega$, and hence we expect the pure-dephasing approximation to 
break down for driven QD systems. 

Nevertheless, based on the above intuition and for comparison, we shall introduce the pure-dephasing approximation to the QD dynamics. 
The equation of motion for the QD density operator in the Schr\"{o}dinger picture is written 
\beq
\frac{\mathrm{d}{\rho_{\rm S}(t)}}{\mathrm{d}t}=-i[H_S,\rho_{\rm S}(t)]+\frac{1}{2}\gamma_{\mathrm{PD}}\big(\sigma_z \rho_{\rm S}(t)\sigma_z-\rho_{\rm S}(t)\big),
\label{rhodotPD}
\eeq
where $\sigma_z=\ketbra{X}{X}-\ketbra{0}{0}$. As we shall see, the pure-dephasing rate $\gamma_{\mathrm{PD}}$ determines how quickly 
off-diagonal elements of $\rho_{\rm S}(t)$ decay. In the crudest approximation $\gamma_{\mathrm{PD}}$ can be taken to 
be some constant whose value is fixed by experimental observations. To go beyond this purely phenomenological treatment, 
the most important features of phonon-induced dephasing (at finite driving $\Omega$) can be captured by the form~\cite{flagg09,McCutcheon2013}
\beq
\gamma_{\mathrm{PD}}=\pi \alpha k_B T \Omega^2,
\label{gammaPD}
\eeq
which will be justified in Section~\ref{WeakCouplingMasterEquation} below. 
Here, $\alpha$ is a measure of the exciton--phonon coupling strength defined through the phonon spectral density, 
$J_{\rm ph}(\omega)$, introduced in Eq.~({\ref{phononSD}}), 
and $T$ is the QD sample temperature. 

In order to solve Eq.~({\ref{rhodotPD}}), it is helpful to introduce the Bloch vector, 
defined as $\vec{\alpha}(t)=(\av{\sigma_x},\av{\sigma_y},\av{\sigma_z})$ with $\av{\sigma_i}=\alpha_i=\mathrm{Tr}_{\rm S}[\sigma_i\rho_{\rm S}(t)]$ 
for $i=\{x,y,z\}$, and from which all expectation values pertaining to the QD degrees of freedom can be calculated. 
For a Markovian master equation, the Bloch vector obeys a differential equation of the form
\beq
\dot{\vec{\alpha}}(t)=M\cdot\vec{\alpha}(t)+\vec{b}.
\label{alphadotGeneral}
\eeq
%
From Eq.~({\ref{rhodotPD}}) we find that in the pure-dephasing approximation 
%
\beq
M_{\mathrm{PD}} = \left(
\begin{array}{ccc}
-\gamma_{\mathrm{PD}} & -\delta & 0 \\
\delta & -\gamma_{\mathrm{PD}} & -\Omega \\
0 & \Omega & 0 \end{array}
\right),
\eeq
and $\vec{b}_{\mathrm{PD}}=(0,0,0)$. Since the inhomogeneous term in Eq.~({\ref{alphadotGeneral}}) is zero, the 
pure-dephasing steady-state solution is the null vector, 
$\vec{\alpha}(\infty)=-M_{\mathrm{PD}}^{-1}\cdot\vec{b}_{\mathrm{PD}}=(0,0,0)$, which corresponds to a maximally mixed 
state $\rho_{\rm S}(\infty)=\frac{1}{2}\openone$, rather than a thermal equilibrium state at temperature $T$. For a QD prepared initially in its ground state we have $\vec{\alpha}(0)=(0,0,-1)$, 
and solving Eq.~({\ref{alphadotGeneral}}) under resonant driving conditions, $\delta=0$, we find $\alpha_x(t)=0$, while 
\begin{align}
\!\!\alpha_y(t) &= \e^{-\gamma_{\mathrm{PD}} t /2} \frac{2 \Omega}{\xi_{\mathrm{PD}}} \sin\big(\xi_{\mathrm{PD}} t/2\big), \\
\!\!\alpha_z(t) &=- e^{-\gamma_{\mathrm{PD}} t /2}\Big[\cos\Big(\frac{\xi_{\mathrm{PD}} t}{2}\Big)
+\frac{\gamma_{\mathrm{PD}}}{\xi_{\mathrm{PD}}}\sin\Big(\frac{\xi_{\mathrm{PD}} t}{2}\Big)\Big],
\end{align}
where $\xi_{\mathrm{PD}}=\sqrt{4 \Omega^2-\gamma_{\mathrm{PD}}^2}$. These solutions demonstrate that within the 
pure-dephasing approximation the QD performs damped Rabi oscillations, which 
in principle become over-damped when $\gamma_{\mathrm{PD}}> 2\Omega$, and relaxes in the long time limit to a maximally mixed state.

\subsection{Weak-coupling master equation}
\label{WeakCouplingMasterEquation}

Moving on from the phenomenological pure-dephasing approximation, we 
now derive equations of motion in a more rigorous manner from the microscopic Hamiltonian of Eq.~({\ref{HTotal}}). Given the general master equation form of Eq.~({\ref{rhodotD2}}) 
found in Section~\ref{TheoreticalBackground}, we simply insert the appropriate quantities from Eqs.~({\ref{HS}})-(\ref{HE}). The 
central approximation made here is that the master equation we derive is valid to second order in the exciton-phonon  
interaction Hamiltonian of Eq.~({\ref{HI}}).

As outlined, Eq.~({\ref{rhodotD2}}) is obtained assuming an interaction Hamiltonian of the form 
$\sum_{i} A_{i}\otimes B_{i}$ with both $A_i$ and $B_i$ Hermitian. Inspection of Eq.~({\ref{HI}}) reveals that in the present 
case we have only a single term in our sum, and we can therefore assign $A_z=\ketbra{X}{X}=\smash{\frac{1}{2}(\openone+\sigma_z)}$ 
and $B_z=\sum_\kv g_\kv(b_\kv^{\dagger}+b_\kv)$. The response of the environment 
is characterised by a single bath correlation function 
$C_{zz}(\tau)=\mathrm{Tr}_\Env[\tilde{B}_z(\tau) \tilde{B}_z(0) \rho_\Env]$. 
Within the interaction picture
\beq
\tilde{B}_z(\tau)=
\sum_\kv g_\kv (b_\kv^{\dagger}\e^{i\w_\kv\tau}+b_\kv \e^{-i\w_\kv \tau}), 
\eeq
and assuming a thermal state for the initial environmental density operator, 
$\rho_\Env(0) = \e^{-\beta H_\Env}/\mathrm{Tr}_\Env[\e^{-\beta H_\Env}]$ with inverse temperature $\beta=1/k_B T$, we obtain the weak-coupling 
correlation function~\cite{ramsay10,ramsay10_2}
\beq
C_{zz}(\tau)=\int_0^{\infty}\mathrm{d}\w J_{\rm ph}(\w)(\cos(\w\tau)\coth(\beta\w/2)-i \sin(\w\tau)), 
\label{CWeak}
\eeq
where we have used results given in Appendix~\ref{CorrelationFunctions} [see Eq.~(\ref{CWeakSingleMode})].  
The system operator Fourier components are found to be 
\begin{align}
A_z(0)=&\frac{1}{2}\openone+\frac{\delta\Omega}{2\eta_{\mathrm{w}}^2}\sigma_x+\frac{\delta^2}{2\eta_{\mathrm{w}}^2}\sigma_z,\\
A_z(\eta_{\mathrm{w}})=&-\frac{\delta\Omega}{4\eta_{\mathrm{w}}^2}\sigma_x+\frac{i\Omega}{4\eta_{\mathrm{w}}}\sigma_y+\frac{\Omega^2}{4\eta_{\mathrm{w}}^2}\sigma_z,
\end{align}
where $\eta_{\mathrm{w}}=\sqrt{\delta^2+\Omega^2}$ is the generalised Rabi frequency, $A_z(-\eta_{\mathrm{w}})=A_z^{\dagger}(\eta_{\mathrm{w}})$, and it can be verified that $\sum_\zeta A_z(\zeta)=\ketbra{X}{X}$. The `$\mathrm{w}$' subscripts used on various quantities here and below 
remind us that they belong to the weak-coupling theory, which will become important when we explore the more sophisticated variational theory in 
Section~{\ref{VariationalTheory}}. 
Putting these expressions into Eq.~({\ref{rhodotD2}}) 
we find that the Bloch vector again obeys an equation of motion 
of the general form given in Eq.~({\ref{alphadotGeneral}}), this time with 
%
\beq
M_\mathrm{W} = \left( \begin{array}{ccc} 
-\sfrac{\Omega^2}{\eta_{\mathrm{w}}^2} \Gamma_\mathrm{w1} & -\delta' & -\sfrac{\delta \Omega}{\eta_{\mathrm{w}}^2}\Gamma_\mathrm{w1} \\
\delta' & - \sfrac{\Omega^2}{\eta_{\mathrm{w}}^2}\Gamma_\mathrm{w1} & - \Omega' \\ 
0 & \Omega & 0 \end{array}
\right),
\label{MW}
\eeq
and $\vec{b}_\mathrm{W}=\big(-\sfrac{\Omega}{\eta_{\mathrm{w}}}\kappa_\mathrm{w1},\sfrac{\delta\Omega}{\eta_{\mathrm{w}}^2}[\lambda_{\mathrm{w2}}-\zeta_\mathrm{w1}],0\big)$. Here we have defined the quantities 
\begin{align}
\label{Gammaw1}
\Gamma_\mathrm{w1}=&\sfrac{1}{4}\big(\gamma_{zz}(\eta_{\mathrm{w}})+\gamma_{zz}(-\eta_{\mathrm{w}})\big),\\
\kappa_\mathrm{w1}=&\sfrac{1}{4}\big(\gamma_{zz}(\eta_{\mathrm{w}})-\gamma_{zz}(-\eta_{\mathrm{w}})\big),\\
\lambda_\mathrm{w1} = &\sfrac{1}{2}\big(S_{zz}(\eta_{\mathrm{w}})-S_{zz}(-\eta_{\mathrm{w}})\big),\\
\zeta_{\mathrm{w1}} = & \sfrac{1}{2}\big(S_{zz}(\eta_{\mathrm{w}})+S_{zz}(-\eta_{\mathrm{w}})\big),\label{zetaw1}\\
\lambda_{\mathrm{w2}}=&S_{zz}(0)\label{lambdaw2},
\end{align}
in terms of the weak-coupling response function, with 
\begin{align}
\Omega'=\Omega+\frac{\Omega}{\eta_{\mathrm{w}}}\lambda_\mathrm{w1},
\end{align}
and the detuning now given by
\begin{align}\label{deltaprime}
\delta' = \delta+\lambda_{\mathrm{w2}}.
\end{align} 
Note that we have used $\gamma_{zz}(0)=0$ in arriving at the coefficient forms given in Eq.~({\ref{MW}}), 
which is valid only in the Markov approximation. 

The rates $\gamma_{zz}(\zeta)$ and energy shifts $S_{zz}(\zeta)$ 
are defined in accordance with 
Eqs.~({\ref{KDefinition}}) and ({\ref{CWeak}}). In the Markovian limit of the weak-coupling theory, 
it is possible with the help of Eq.~(\ref{diracreln}) to obtain certain analytic expressions. Taking the one-sided Fourier transform of 
the weak-coupling correlation function 
we find 
\begin{align}
\gamma_{zz}(\w)=&\;2\pi\int_0^{\infty} \mathrm{d}\w' J_{\rm ph}(\w')\times\nonumber\\
&\left(\delta(\w+\w')N(\w')+\delta(\w-\w')[N(\w')+1]\right),
\end{align}
and 
\begin{align}
S_{zz}(\w)=P\int_0^{\infty} \mathrm{d}\w' J_{\rm ph}(\w')\left(\frac{N(\w')}{\w+\w'}+\frac{N(\w')+1}{\w-\w'}\right),
\end{align}
where $N(\w)=(\e^{\beta\w}-1)^{-1}$ is the thermal occupation number. 
With these expressions we find the weak-coupling 
rate can be written 
\beq
\Gamma_\mathrm{w1} = \frac{\pi}{2} J_{\rm ph}(\eta_{\mathrm{w}})\coth(\beta\eta_{\mathrm{w}}/2),
\label{GammaW}
\eeq
while 
\begin{align}
\lambda_\mathrm{w1} = &{P}\int_0^{\infty}\mathrm{d}\w J_{\rm ph}(\w)\frac{\eta_{\mathrm{w}}\coth(\beta\w/2)}{\eta_{\mathrm{w}}^2-\w^2},\\
\zeta_\mathrm{w1} = &{P}\int_0^{\infty}\mathrm{d}\w J_{\rm ph}(\w)\frac{\w}{\eta_{\mathrm{w}}^2-\w^2},
\end{align}
and 
\begin{align}\label{polshift}
\lambda_\mathrm{w2}=-\int_0^{\infty}\mathrm{d}\w \frac{J_{\rm ph}(\w)}{\w}, 
\end{align}
the last of which we shall refer to as the polaron shift, as it is responsible for an environmentally induced redefinition of the QD resonance conditions, see Eq.~(\ref{deltaprime}).
This will also be an important quantity when we consider the polaron and variational 
theories in subsequent sections, where its origins will become clearer. 

With the weak-coupling Bloch equations 
having been found, 
we can now assess the validity of the pure-dephasing approximation introduced in Section~{\ref{PureDephasingApproximation}}. 
For driving at the phonon-shifted resonance, $\delta=\int_0^{\infty}\mathrm{d}\w J_{\rm ph}(\w)/\w$ 
(i.e.~$\delta'=0$), we can solve the Bloch equations for the QD population difference $\alpha_z$ 
assuming the dot to be initialised in its ground state, $\alpha_z(0)=-1$. This gives 
\begin{align}
\alpha_z(t)=q - (q+1)\e^{-\gamma_{\mathrm{w}}t/2}
\Big[\cos\Big(\frac{\xi_{\mathrm{w}} t}{2}\Big)+\frac{\gamma_{\mathrm{w}}}{\xi_{\mathrm{w}}}\sin\Big(\frac{\xi_{\mathrm{w}} t}{2}\Big)\Big],
\label{alphazWC}
\end{align}
where the population damping rate is $\gamma_{\mathrm{w}}=(\Omega/\eta_{\mathrm{w}})^2\Gamma_{\mathrm{w1}}$, the generalised Rabi frequency becomes 
$\xi_{\mathrm{w}}=\sqrt{4\Omega\Omega'-\gamma_\weak^2}$,  
and we have defined the quantity 
$q=(\delta\Omega/\eta_\weak^2)[\lambda_{\mathrm{w2}}-\zeta_{\mathrm{w1}}]/\Omega'$. For 
most parameters of interest it can be shown numerically that $q\ll 1$, and we therefore neglect it 
in the discussion that follows. Now, from 
Eq.~({\ref{GammaW}}), we see that if $\delta\ll\eta_\weak\ll k_B T,\w_c$, the rate appearing 
in the weak-coupling solution can be approximated as 
\beq
\gamma_\weak \approx \pi \alpha k_B T \Omega^2,
\eeq
which is precisely the pure-dephasing rate introduced in Eq.~({\ref{gammaPD}}). 
Furthermore, if we replace the driving strength in the pure-dephasing approximation with 
the phonon renormalised version, 
i.e. we let $\Omega^2\to\Omega\Omega'$, then as far as the QD populations are concerned, 
the pure-dephasing approximation becomes equivalent to the weak-coupling theory.

Important differences, however, are present in the dynamics of the QD coherences, particularly in 
the expectation value $\alpha_x$. While in the pure-dephasing case we 
have simply $\dot{\alpha}_x=-\gamma_{\mathrm{PD}}\alpha_x$, 
from Eq.~({\ref{MW}}) 
the corresponding equation 
of motion in the weak-coupling theory is considerably more complicated. Notably, in regimes where $\Omega\gg\delta$, 
we have $\dot{\alpha}_x\approx-\gamma_{\mathrm{w}}\alpha_x-\kappa_{\mathrm{w1}}$ with 
%
\beq
\kappa_\mathrm{w1}=\frac{\pi}{2} J_{\rm ph}(\Omega), 
\eeq
%
leading to the steady-state solution 
$\alpha_x(\infty)=-\kappa_\mathrm{w1}/\gamma_{\mathrm{w}}=-\tanh(\beta\Omega/2)$, 
which is the expected expression in thermal equilibrium and can be significant for low temperatures and/or large driving strengths. 
This observation demonstrates a serious deficiency of the pure-dephasing approximation, which 
instead gives $\alpha_x(\infty)=0$, and thus essentially corresponds to a high temperature limit. For driven systems it does not, therefore, predict the correct behaviour of the QD coherences outside the semiclassical regime~\footnote{Even in the undriven case ($\Omega\rightarrow0$), where pure-dephasing is the appropriate description, Eq.~(\ref{rhodotPD}) is overly simplistic, as it cannot describe the non-Markovian dynamics that arises due to polaron formation, see Section~\ref{IBM}.}. 
More generally, it can be said that 
the pure-dephasing approximation does not obey detailed balance conditions, and therefore does not 
lead to the correct (thermal) steady-state. If we are only interested in the QD populations, this discrepancy may be  
unimportant, since on resonance at least, the pure-dephasing approximation does indeed predict the correct qualitative 
behaviour for $\alpha_z$. 
Even in this case, however, it is somewhat phenomenological, and the dependence of the dephasing rate (and any environment induced frequency shifts) on the system parameters must be incorporated by hand. 
Moreover, as we shall see in Section~{\ref{Emission}}, the fact that phonons lead to behaviour for the coherence $\alpha_x$ which differs from the pure-dephasing approximation can be vitally important 
when the QD emission characteristics are considered.

\begin{figure}[t]
\begin{center}
\includegraphics[width=0.49\textwidth]{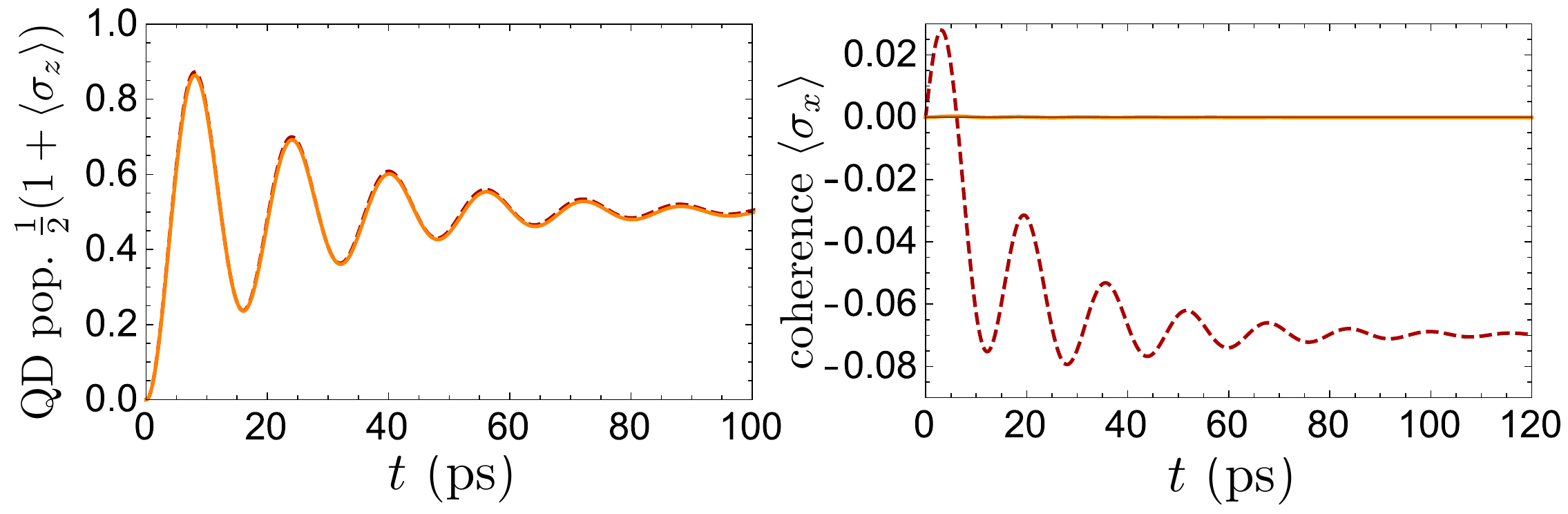}
\caption{QD excited state population (left) and coherence (right) as a function of 
time, calculated within the pure-dephasing approximation (solid, orange curves), and 
with the weak-coupling theory (dashed, red curves). We see that while the QD populations 
are closely matching, the coherences differ considerably.  
Parameters used: $\Omega=0.5~\mathrm{ps}^{-1}$, 
$T=30~\mathrm{K}$, 
$\alpha=0.027~\mathrm{ps}^2$ and $\w_c=2.2~\mathrm{ps}^{-1}$, 
and we drive the QD at the polaron-shifted resonance, $\delta=\int_0^{\infty}\mathrm{d}\w J_{\rm ph}(\w)/\w$ 
(we have used 
$\Omega\to\sqrt{\Omega\Omega'}$ and $\delta=0$ in the 
pure-dephasing theory, see main text).}
\label{PDPlots}
\end{center}
\end{figure}

To illustrate these points, in Fig.~\ref{PDPlots} we show the excited state 
population and coherence of a driven QD calculated within the pure-dephasing approximation 
(solid, orange curves) and with the weak-coupling theory (dashed, red curves). As discussed above, 
in order to achieve the best possible comparison, in the weak-coupling theory we have set $\Omega=0.5~\mathrm{ps}^{-1}$ and 
$\delta=-\lambda_{\mathrm{w2}}=-S_{zz}(0)=\int_0^{\infty}\mathrm{d}\w J_{\rm ph}(\w)/\w$, though in the pure-dephasing approximation we use 
$\delta=0$ and $\Omega\to\sqrt{\Omega(\Omega+\lambda_{\mathrm{w1}})}$.
Once these replacements have been made 
we see, as expected, that both theories predict the same behaviour for 
the QD population, with the damping rate and oscillation period matching well. However, 
for these experimentally relevant parameters~\cite{ramsay10,ramsay10_2}, 
the pure-dephasing approximation does not predict 
the correct evolution for the QD coherence, either in the transient regime or the steady-state. 

\begin{figure}[t]
\begin{center}
\includegraphics[width=0.49\textwidth]{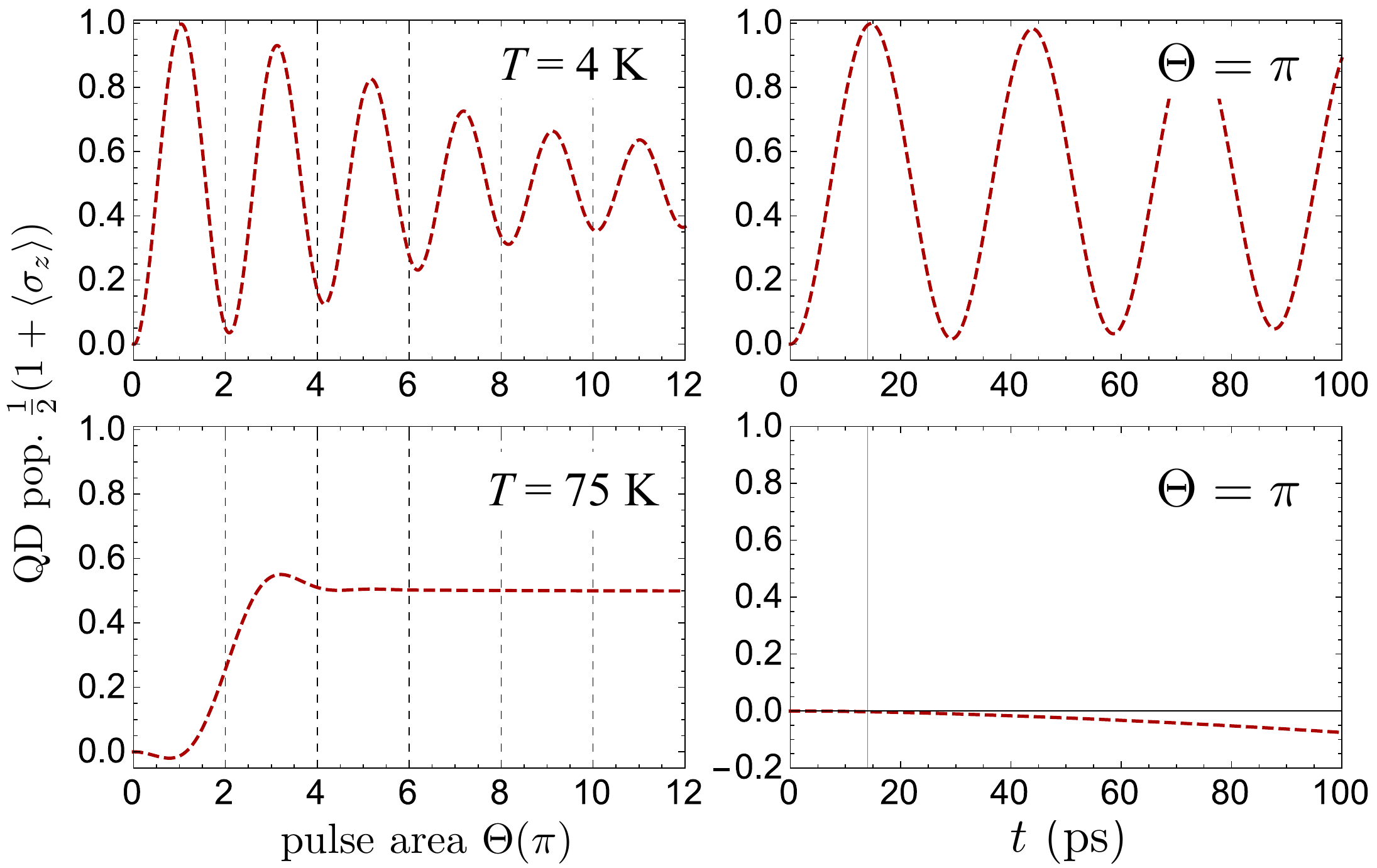}
\caption{QD excited state population as a function of pulse area (left) and 
time (right), calculated using the weak-coupling theory. 
The theory predicts increased damping with larger pulse area (equivalently, 
stronger driving). At elevated temperatures the theory becomes unphysical, as 
can be seen by the prediction of negative populations for $T=75~\mathrm{K}$. Other parameters are $\alpha=0.027~\mathrm{ps}^2$ and $\w_c=2.2~\mathrm{ps}^{-1}$.}
\label{WCPlots}
\end{center}
\end{figure}

Let us now examine the weak-coupling theory itself in more detail. In addition to time domain dynamics, 
we may also consider the QD excited state population as a function of {\emph{pulse area}} $\Theta=\Omega\Delta\tau$, 
which we calculate by evolving the QD density operator for a fixed duration $\Delta\tau$ and 
increasing the Rabi frequency $\Omega$. In the left panels of Fig.~\ref{WCPlots} we show 
the QD population as a function of pulse area for fixed $\Delta\tau=14~\mathrm{ps}$ at the two different 
temperatures indicated. The right panels display the corresponding time domain dynamics for 
a representative pulse area of $\Theta=\pi$ (corresponding to a pulse strength of $\Omega=\pi/14~\mathrm{ps}^{-1}$). 
As before, we drive the QD at its polaron shifted transition frequency, 
and use material parameters relevant to the experiments in Refs.~\cite{ramsay10,ramsay10_2}.

For the plots corresponding to $T=4~\mathrm{K}$ we see that the weak-coupling theory predicts 
increased damping with larger pulse area $\Theta$ (equivalently Rabi frequency $\Omega=\Theta/\Delta\tau$). This 
can be understood from Eqs.~({\ref{GammaW}}) and ({\ref{alphazWC}}) where 
the damping rate in the weak-coupling theory is found to be proportional to the spectral density evaluated at 
$\eta_{\mathrm{w}}=\sqrt{\Omega^2+\delta^2}$, and provided 
$\delta\ll\Omega\ll k_B T,\w_c$, we have simply 
$\gamma_{\mathrm{w}}\approx \alpha \pi k_B T \Omega^2$. We therefore expect the damping rate to increase linearly with both temperature and 
the square of the Rabi frequency in this regime, which has also been confirmed experimentally~\cite{ramsay10,ramsay10_2,ulrich11_short,Wei2014}. 
At elevated temperatures we find that the weak-coupling theory begins to break down, and 
when $T=75~\mathrm{K}$ and $\Theta\sim\pi$ it predicts an unphysical negative excited state 
population. As was shown in Ref.~\cite{mccutcheon10_2}, this failure arises 
due to an misestimation of the Rabi frequency renormalisation captured through $\Omega'$, and ultimately indicates  
that the weak system-environment coupling assumption no longer holds. More specifically, being perturbative in the exciton-phonon interaction term, the weak-coupling theory captures only single-phonon processes. At elevated temperatures (or larger $\alpha$), multiphonon processes can play a significant role in the QD dynamics, in which case we must look for alternative methods to describe our open QD system. 


\section{QD Dynamics - Polaron Theory}
\label{PolaronTheory}

Though the weak-coupling theory presented in Section~\ref{WeakCoupling} works well in the limit of 
small exciton-phonon coupling strengths and/or low temperatures, as we have seen, outside these limits it becomes invalid 
and in certain parameter regimes can even predict unphysical 
behaviour. To go beyond weak-coupling it is necessary to formulate an alternative perturbative expansion, and in this section 
we shall introduce one way of doing this, namely polaron theory. The essence of this approach is to apply a (physically motivated) unitary polaron transformation to the complete Hamiltonian~\cite{mahan,hohenester07}, 
and then derive a master equation to second order in the interaction terms in the transformed basis~\cite{mccutcheon10_2}. As we shall see, the polaron  
basis is such that we are able to incorporate much of the effect of the original system-environment interaction into our free Hamiltonian, which 
we then treat to all orders. Furthermore, we shall show that in contrast to the weak-coupling approach, 
the polaron master equation remains valid for arbitrary coupling strength and temperature (broadly speaking) provided that the QD Rabi
frequency is smaller than the 
phonon environment cut-off~\cite{mccutcheon10_2}. This condition is often more 
easily 
fulfilled experimentally than those required for the weak-coupling approximation to hold.
For example, in 
current relevant 
experiments, 
continuous-wave driving strengths typically give rise to Rabi frequencies 
up to around $200~\mu\mathrm{eV}\approx 20\times 2 \pi~\mathrm{GHz}$ 
at most~\cite{ulrich11_short,Wei2014}, 
whereas for QDs on the $\sim 10~\mathrm{nm}$ scale phonon cut-off frequencies are of the 
order of $1$~meV~\cite{nazir08,ramsay10_2}.

\subsection{Independent boson model}
\label{IBM}

To motivate the polaron approach and understand its physical meaning, it is instructive to first consider the zero driving limit of our 
complete Hamiltonian in Eq.~({\ref{HTotal}}). In 
this case $\Omega\to0$, $\delta\to \epsilon_X$, and the Hamiltonian becomes what is known as the independent 
boson model, which permits an exact solution~\cite{mahan,krummheuer02,hohenester07}. 
Specifically, we have $H\to H_{\mathrm{IB}}$ where 
\beq
H_{\mathrm{IB}}=\epsilon_X\ketbra{X}{X}+H_\II+H_\Env,
\eeq
with $H_\II=\ketbra{X}{X}\sum_\kv g_\kv (b_\kv^{\dagger}+b_\kv)$ and $H_\Env=\sum_\kv \w_\kv b_\kv^{\dagger}b_\kv$ as before. 
We now consider the action of the unitary polaron transformation, defined as 
\beq
H_{\mathrm{IB,P}}=\e^S H_{\mathrm{IB}} \e^{-S},
\label{HIBP}
\eeq
with $S=\ketbra{X}{X}\sum_\kv h_\kv(b_\kv^{\dagger}-b_\kv)$ and $h_\kv=g_\kv/\w_\kv$, such that we can write 
\beq
\e^{\pm S}=\ketbra{0}{0}+\ketbra{X}{X} B_{\pm}.
\label{eS}
\eeq
Here, $B_{\pm}=\prod_\kv D_\kv(\pm h_\kv)$ are defined in terms of displacement operators  
$D_\kv(\pm h_\kv)=\exp[\pm (h_\kv b_\kv^{\dagger}- h_\kv^* b_\kv)]$ 
whose properties we discuss in Appendix~{\ref{coherentstates}}. 
The transformed Hamiltonian takes the uncoupled form~\cite{hohenester07}
\beq
H_{\mathrm{IB,P}}=\epsilon_X'\ketbra{X}{X}+H_\Env,
\eeq
where $\epsilon_X'=\epsilon_X - \sum_\kv g_\kv^2/\w_\kv$ is the phonon-shifted QD transition frequency. 
In the continuum limit, $\sum_\kv g_\kv^2/\w_\kv\rightarrow\int_0^{\infty}{\mathrm d}\omega J_{\rm ph}(\omega)/\omega$, and we now see why we termed $\lambda_{\rm w2}$ in Eq.~(\ref{polshift}) the polaron shift; it is the displacement energy associated with formation of a polaron (charge-phonon) quasiparticle due to the exciton-phonon interaction. 

To calculate the dynamics of the reduced density operator describing the QD degrees of freedom we 
use Eq.~({\ref{HIBP}}) to write the time evolution operator as 
\beq
U(t)=\e^{-i H_{\mathrm{IB}} t}=\e^{-S} \e^{-i H_{\mathrm{IB,P}} t} \e^{S},
\eeq
or equivalently
\beq
U(t)=\ketbra{0}{0}U_\Env(t)+\e^{- i \epsilon_X' t}\ketbra{X}{X}B_- U_\Env(t) B_+,
\eeq
with $U_\Env(t)=\e^{-i H_\Env t}$. 
Thus $\rho_\Sys(t)=\mathrm{Tr}_\Env\big[U(t)\rho(0) U^{\dagger}(t)\big]$, and assuming 
a factorising initial state $\rho(0)=\rho_\Sys(0)\rho_\Env(0)$, we find 
\begin{align}
\rho_\Sys(t)=\;\;&\rho_{00}\ketbra{0}{0}+\rho_{XX}\ketbra{X}{X} \vphantom{\e^{\epsilon'}} \nonumber\\
&+\rho_{0X}\ketbra{0}{X}\e^{i\epsilon_X' t} C_{-+}^*(t) \nonumber\\
&+\rho_{X0}\ketbra{X}{0}\e^{-i\epsilon_X' t} C_{-+}(t),
\label{rhoIBM}
\end{align} 
where $\rho_{ij}=\bra{i} \rho_\Sys(0)\ket{j}$ for $i,j=\{0,X\}$. Here we have defined the environment 
correlation function using our usual notation
\beq
C_{-+}(t)=\mathrm{Tr}_\Env\left[ \tilde{B}_-(t)B_+ \rho_\Env(0)\right],
\eeq
with $\tilde{B}_{\pm}(t)=U_\Env^{\dagger}(t)B_{\pm} U_\Env(t)=\prod_\kv D_\kv(\pm h_\kv \e^{i \w_\kv t})$. 
We perform the trace using the coherent state representation, as 
detailed in Appendix~{\ref{CorrelationFunctions}} [see Eq.~(\ref{Bpmappendix})], 
and find that for an 
initial thermal state of the phonon environment the correlation function becomes 
$C_{-+}(t)=\av{B}^2 \e^{\phi(t)}$, where 
\beq
\phi(t)=\int_0^{\infty}\mathrm{d}\w \frac{J_{\rm ph}(\w)}{\w^2}\Big(\cos\w t \coth(\beta\w/2)-i\sin\w t\Big),
\label{PhononPropagator}
\eeq
is the phonon propagator and 
\begin{align}
\av{B}&={\rm Tr}_{\rm E}[B_{\pm}\rho_{\rm E}(0)]\nonumber\\
&=\exp\left[-\frac{1}{2}\int_0^{\infty}\mathrm{d}\w\frac{J_{\rm ph}(\w)}{\w^2}\coth(\beta\w/2)\right],
\label{Bav}
\end{align}
which can be written $\av{B}=\e^{-\frac{1}{2}\phi(0)}$. 
The factor $\av{B}$ plays an important role in the polaron master equation to be derived below, as it is responsible for a phonon-induced renormalisation of the QD Rabi frequency. 

\begin{figure}[t]
\begin{center}
\includegraphics[width=0.49\textwidth]{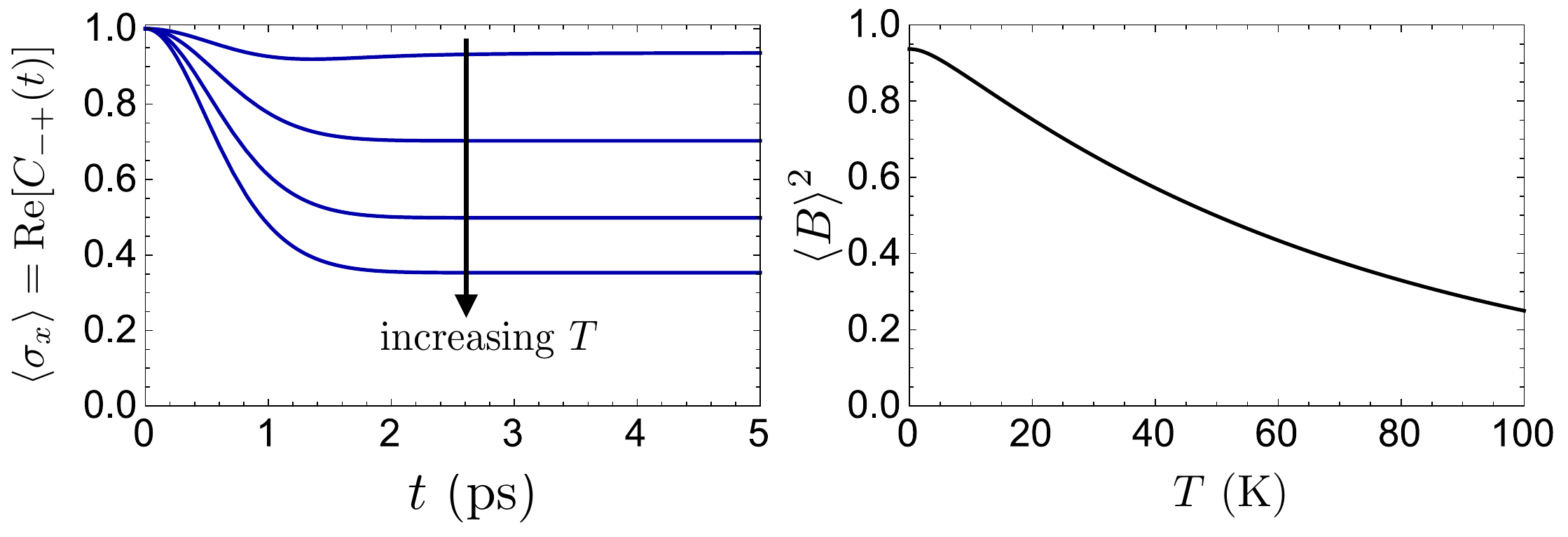}
\caption{Left: Evolution of the QD coherence as a function of time calculated within the independent boson model (zero driving limit). 
The curves are plotted for temperatures $T=0,25,50,75~\mathrm{K}$ ordered as indicated. Right: Steady-state 
values are given by $\av{B}^2$ which we plot as a function of temperature. 
QD parameters used: $\alpha=0.027~\mathrm{ps}^2$ and $\w_c=2.2~\mathrm{ps}^{-1}$.}
\label{IBMPlots}
\end{center}
\end{figure}

Returning to Eq.~({\ref{rhoIBM}}), we can see immediately that within the independent boson model 
the QD populations are stationary, while the coherences contained in the off-diagonal elements 
evolve in some way, as expected for a pure-dephasing process. If we consider a QD initialised in an equal superposition of its ground 
and excited state, we have $\rho_{ij}(0)={1}/{2}$ for all $i,j=\{0,X\}$. Calculating the subsequent evolution of the coherence 
$\av{\sigma_x}=\mathrm{Tr}_\Sys[\rho_\Sys(t)\sigma_x]$ we find 
\begin{align}
\av{\sigma_x}=\mathrm{Re}[C_{-+}(t)],
\end{align}
where we have moved into a frame rotating 
at the phonon-shifted QD transition frequency $\epsilon_X'$. 
In Fig.~{\ref{IBMPlots}} we plot the dynamics of this coherence for increasing temperature 
as indicated. 
In all cases we see an initial rapid decay on a picosecond timescale followed by a plateau to a constant temperature dependent value,  
$\lim_{t\to\infty}C_{-+}(t) = \av{B}^2$, which we plot 
on the right. Physically, we are seeing the effects of polaron formation~\cite{hohenester07}. As the phonon bath relaxes from its initial thermal state to a displaced thermal state due to the exciton-phonon interaction term, coherence is lost from the QD excitonic degrees of freedom. However, once the displaced equilibrium is reached (and the polaron formed), no further loss of coherence is observed. Note that though this process is a pure-dephasing one, it cannot be modelled by a simple Markovian rate. 

\subsection{Master equation in the polaron frame}
\label{PolaronMasterEquation}

As we have just demonstrated, 
in the zero driving limit our 
coupled QD-phonon Hamiltonian may be exactly diagonalised by a unitary polaron transformation. This motivates us 
to suppose that outside the zero driving limit, the polaron transformation should approximately diagonalise 
our Hamiltonian, at least to the level of describing the process of bath relaxation and polaron formation. The change of basis will then leave some residual interaction term to which we can apply perturbation theory.

To derive the QD master equation in the polaron frame, we therefore return to the full Hamiltonian in 
Eq.~({\ref{HTotal}}) and apply the same unitary transformation defined by~\cite{wurger98,hohenester07,mccutcheon10_2,roy11}
\beq
H_\Pol = \e^S H \e^{-S},
\label{HP0}
\eeq
where $S=\ketbra{X}{X}\sum_\kv h_\kv(b_\kv^{\dagger}-b_\kv)$ with $h_\kv=g_\kv/\w_\kv$, as in Eq.~({\ref{HIBP}}). 
After the transformation we now find 
\beq
H_\Pol= \delta'\ketbra{X}{X}+\frac{\Omega}{2}\big(\ketbra{0}{X}B_- +\ketbra{X}{0}B_+\big)+H_\Env,
\label{HP1}
\eeq
where $\delta' = \delta-\sum_\kv g_\kv^2/\w_\kv$ is again the polaron shifted detuning. As anticipated, although the original 
exciton-phonon coupling term has been removed, the second term in Eq.~({\ref{HP1}}) 
represents a new interaction, involving excitation and de-excitation of the QD along with the appropriate environmental displacement captured through  
$B_{\pm}=\prod_\kv D_\kv(\pm h_\kv)$. 
At this stage it may seem reasonable to 
identify this term as the system-environment interaction and attempt to derive a master equation taking 
it as a perturbation. Note, however, that it has a non-zero expectation 
value with respect to an environmental thermal state $\rho_{\rm E}$, i.e.~$\mathrm{Tr}_\Env[(\ketbra{0}{X}B_- +\ketbra{X}{0}B_+)\rho_\Env]=\av{B}\sigma_x$. 
It is therefore more appropriate 
to define the system-environment interaction with reference to this 
expectation, as we shall then be performing a perturbation expansion in fluctuations around the thermal average. To do so 
we simply add $\av{B}\sigma_x$ to the part we define as the system Hamiltonian while 
subtracting it from the interaction, leaving the total Hamiltonian unchanged.
We thus write
\beq
H_\Pol = H_{\mathrm{SP}}+H_{\mathrm{IP}}+H_\Env,
\label{HP2}
\eeq
where 
\beq
H_{\mathrm{SP}}=\delta'\ketbra{X}{X}+\frac{\Omega_\mathrm{p}}{2}\sigma_x,
\label{HSP}
\eeq
with $\Omega_\mathrm{p}=\Omega \av{B}$ the bath renormalised driving strength. The interaction term becomes
\beq\label{HIPav}
H_{\mathrm{IP}}=\frac{\Omega}{2}\Big(\!\ketbra{0}{X}[B_--\av{B}]+\ketbra{X}{0}[B_+-\av{B}]\Big),
\eeq
which we put in the more convenient form $H_{\mathrm{IP}}=(\Omega/2)(\sigma_x B_x+\sigma_y B_y)$, with 
%
\begin{align}
B_x&=\frac{1}{2}\big(B_++B_--2\av{B}),\nonumber\\
B_y&=\frac{1}{2i}\big(B_--B_+).\label{BxBy}
\end{align}
We note that no approximations have been made in arriving at Eqs.~({\ref{HP2}})-(\ref{HIPav}) from Eq.~({\ref{HP1}}). It is simply a 
matter of redefining what are referred to as our system and interaction Hamiltonians. Importantly though, 
we now have an explicit renormalised driving term in the system Hamiltonian, containing all orders 
of the original system-environment interaction. This renormalisation results from polaron formation, with the classical field now driving transitions not between the original QD exciton states, but between states associated also with the relevant phonon displacements. 

We are now in a position to calculate the correlation functions that enter the polaron frame master equation. 
The form of $H_{\mathrm{IP}}$ tells us that we have two terms in the summations in Eq.~({\ref{rhodotD2}}), i.e.~$i,j=\{x,y\}$ corresponding to the system operators $A_x=(\Omega/2)\sigma_x$ and $A_y=(\Omega/2)\sigma_y$, and 
in principle four correlation functions $C_{ij}(\tau)=\mathrm{Tr}_\Env[\tilde{B}_i(\tau)B_j \rho_\Env]$ corresponding to the 
bath operators in Eq.~({\ref{BxBy}}). Using results outlined in Appendix~\ref{CorrelationFunctions} [see Eqs.~(\ref{Bpmappendix}) and (\ref{Bppappendix})] we find 
\begin{align}
C_{xx}(\tau)&=\frac{\av{B}^2}{2}\big(\e^{\phi(\tau)}+\e^{-\phi(\tau)}-2\big),\label{Cxx}\\
C_{yy}(\tau)&=\frac{\av{B}^2}{2}\big(\e^{\phi(\tau)}-\e^{-\phi(\tau)}\big),\label{Cyy}
\end{align}
with $C_{xy}(\tau)=C_{yx}(\tau)=0$, and $\phi(\tau)$ is again the phonon propagator defined in Eq.~({\ref{PhononPropagator}}). 
The final ingredients needed to construct the explicit form of the polaron master equation are the Fourier components 
of the system operators $A_x$ and $A_y$. These can be calculated straightforwardly from the general definition in 
Eq.~({\ref{fourieroperator}}), however, due to their somewhat cumbersome nature we do not explicitly give them here. Instead, we 
simply note that in analogy to the weak-coupling case, each system operator has in general three Fourier components corresponding 
to the possible energy eigenvalue differences of the system Hamiltonian, 
given in the present case by Eq.~({\ref{HSP}}).

Before we proceed it is important to note that having transformed the total Hamiltonian into the polaron representation, 
our density operator equations of motion are also defined within this frame. 
Referring to our general master equation form in Eq.~({\ref{rhodotD2}}), the density operator appearing 
should now be thought of as the reduced density operator in the polaron basis, defined as 
$\rho_\mathrm{Sp}=\mathrm{Tr}_\Env[ \rho_\mathrm{p}]$ where $\rho_\mathrm{p}=\e^S \rho \e^{-S}$, with 
$\rho$ the total density operator in the original or `lab' frame. As such, when 
we take expectation values of the Pauli operators to 
construct the Bloch vector, 
it too is in the polaron frame, i.e.~we have $\vec{\alpha}_{\mathrm{p}}=(\av{\sigma_x}_\mathrm{p},\av{\sigma_y}_\mathrm{p},\av{\sigma_z}_\mathrm{p})$ 
with $\av{\sigma_i}_\mathrm{p}=\mathrm{Tr}_\Sys[\rho_\mathrm{Sp}\sigma_i]$. To see how these quantities 
are related to those in the original frame we write 
$\av{\sigma_i}=\mathrm{Tr}[\sigma_i \rho]=\mathrm{Tr}[\e^S \sigma_i \e^{-S} \rho_{\mathrm{p}}]$. 
Consistent with our second-order master equation approach, we apply the Born approximation in the polaron frame, 
$\rho_{\mathrm{p}}=\rho_{\mathrm{Sp}}\rho_\Env$, to find 
$\av{\sigma_z}=\av{\sigma_z}_{\mathrm{p}}$, whereas 
$\av{\sigma_x}=\av{B}\av{\sigma_x}_{\mathrm{p}}$ and 
$\av{\sigma_y}=\av{B}\av{\sigma_y}_{\mathrm{p}}$. Thus, while  
the QD population dynamics calculated in the polaron frame is the same as that in the original frame, 
the coherences carry an extra factor of $\av{B}$ within the Born approximation. Of course, on transforming back to the original, lab frame, the total system-environment density operator $\rho$ is no longer necessarily separable due to the polaron-like correlations encoded within our treatment.  

Working through the algebra, 
we therefore 
find that the polaron frame Bloch vector obeys an equation of motion of the form 
$\dot{\vec{\alpha}}_{\mathrm{p}}=M_\Pol\cdot\vec{\alpha}_{\mathrm{p}}+\vec{b}_\Pol$, 
with coefficients given by 
\begingroup
\renewcommand*{\arraystretch}{1.5}
\begin{widetext}
\beq
M_\Pol = \left( \begin{array}{ccc} 
-\Gamma_{\mathrm{p1}} & 
-\big[\delta' +\sfrac{\delta'}{\eta_{\mathrm{p}}}\lambda_{\mathrm{p1}}\big] & 
0 \\
\big[\delta'+\sfrac{\delta'}{\eta_{\mathrm{p}}}\lambda_{\mathrm{p2}}\big] & 
-\big[\sfrac{\Omega_\mathrm{p}^2}{\eta_{\mathrm{p}}^2}\Gamma_{\mathrm{p3}}+\sfrac{\delta'^2}{\eta_{\mathrm{p}}^2}\Gamma_{\mathrm{p2}}\big] & 
-\Omega_\mathrm{p} \\
\sfrac{\delta'\Omega_\mathrm{p}}{\eta_{\mathrm{p}}^2}\big[\Gamma_{\mathrm{p3}}-\Gamma_{\mathrm{p2}}\big] & 
\Omega_\mathrm{p}+\sfrac{\Omega_\mathrm{p}}{\eta_{\mathrm{p}}}\lambda_{\mathrm{p1}} & 
-\big[\sfrac{\Omega_\mathrm{p}^2}{\eta_{\mathrm{p}}^2}\Gamma_{\mathrm{p3}}+\sfrac{\delta'^2}{\eta_{\mathrm{p}}^2}\Gamma_{\mathrm{p2}}+\Gamma_{\mathrm{p1}}\big] \end{array}
\right),
\eeq
\end{widetext}
\endgroup
where we have defined the quantities 
\begin{align}
\label{Gammap1}
\Gamma_{\mathrm{p1}}&=\sfrac{\Omega^2}{4}\big(\gamma_{yy}(\eta_{\mathrm{p}})+\gamma_{yy}(-\eta_{\mathrm{p}})\big),\\
\Gamma_{\mathrm{p2}}&=\sfrac{\Omega^2}{4}\big(\gamma_{xx}(\eta_{\mathrm{p}})+\gamma_{xx}(-\eta_{\mathrm{p}})\big),\\
\Gamma_{\mathrm{p3}}&=\sfrac{\Omega^2}{2}\gamma_{xx}(0),\\
\lambda_{\mathrm{p1}}&=\sfrac{\Omega^2}{2}\big(S_{yy}(\eta_{\mathrm{p}})-S_{yy}(-\eta_{\mathrm{p}})\big),\\
\lambda_{\mathrm{p2}}&=\sfrac{\Omega^2}{2}\big(S_{xx}(\eta_{\mathrm{p}})-S_{xx}(-\eta_{\mathrm{p}})\big),\\
\lambda_{\mathrm{p3}}&=\Omega^2S_{xx}(0),
\end{align}
while 
$\vec{b}_\Pol=\big(-\sfrac{\Omega_\mathrm{p}}{\eta_{\mathrm{p}}}\kappa_{\mathrm{p1}},-\sfrac{\delta'\Omega_\mathrm{p}}{\eta_{\mathrm{p}}^2}[\lambda_\mathrm{p3}-\zeta_\mathrm{p1}],-\sfrac{\delta'}{\eta_{\mathrm{p}}}[\kappa_{\mathrm{p1}}+\kappa_{\mathrm{p2}}]\big)$, 
with 
\begin{align}
\kappa_{\mathrm{p1}}&=\sfrac{\Omega^2}{4}\big(\gamma_{yy}(\eta_{\mathrm{p}})-\gamma_{yy}(-\eta_{\mathrm{p}})\big),\\
\kappa_{\mathrm{p2}}&=\sfrac{\Omega^2}{4}\big(\gamma_{xx}(\eta_{\mathrm{p}})-\gamma_{xx}(-\eta_{\mathrm{p}})\big),\\
\zeta_{\mathrm{p1}}&=\sfrac{\Omega^2}{2}\big(S_{xx}(\eta_{\mathrm{p}})+S_{xx}(-\eta_{\mathrm{p}})),
\label{zetap1}
\end{align}
and $\eta_{\mathrm{p}}=\sqrt{\delta'^2+\Omega_{\mathrm{p}}^2}$. 

\begin{figure}
\begin{center}
\includegraphics[width=0.49\textwidth]{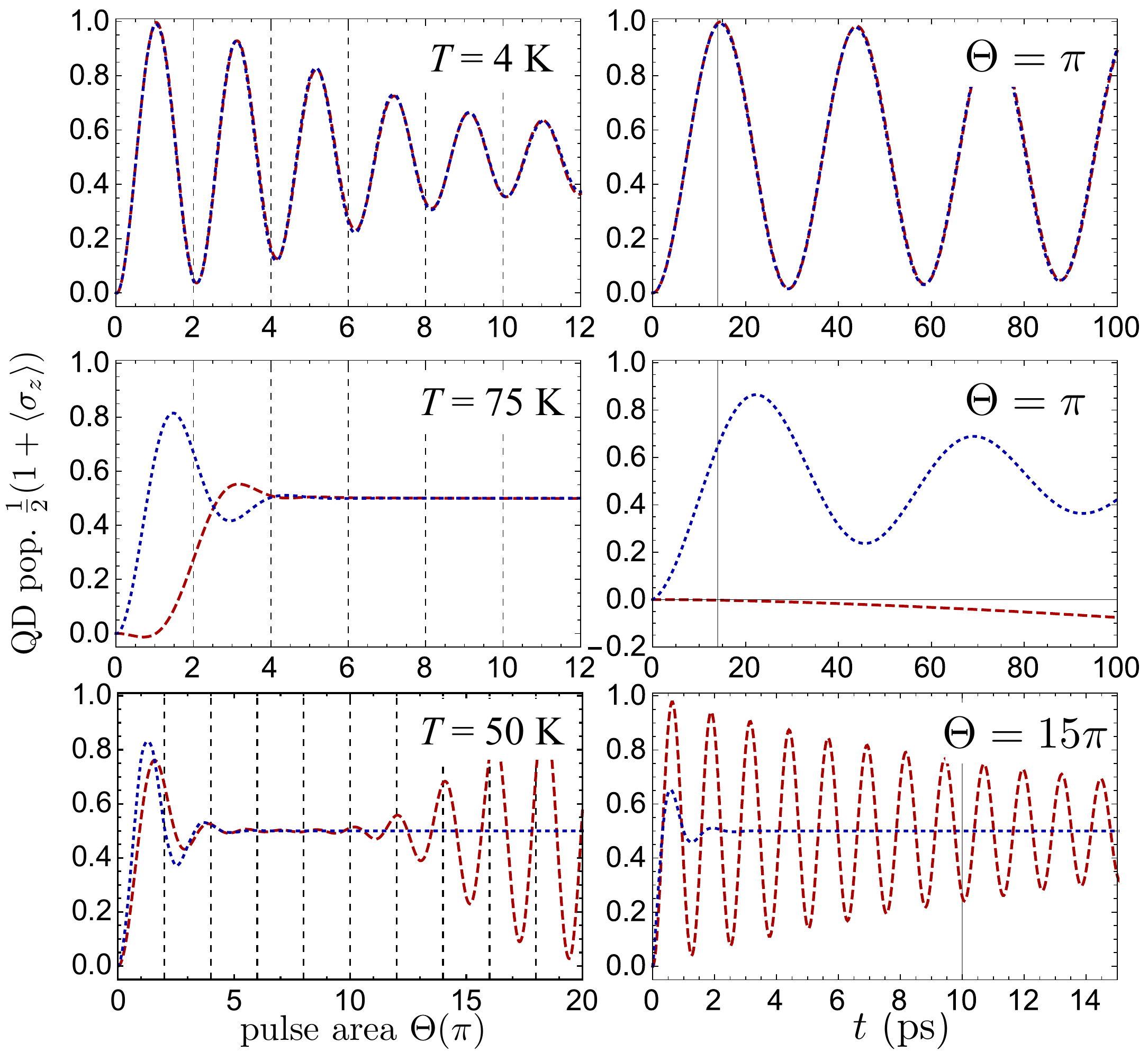}
\caption{QD excited state population as a function of pulse area (left) and 
time (right), calculated using the polaron theory (dotted, blue curves) and the 
weak-coupling theory (dashed, red curves). The first two rows 
correspond to a pulse duration of $\Delta\tau=14~\mathrm{ps}$, while 
the last corresponds to $\Delta\tau=10~\mathrm{ps}$. 
For $T=75~\mathrm{K}$ the 
weak-coupling theory breaks down though the polaron theory remains valid. 
At very large pulse areas an undamped regime is reached, which the polaron theory fails 
to capture (see bottom row).}
\label{PolPlots}
\end{center}
\end{figure}

As in Section~\ref{WeakCoupling},  
we shall now illustrate the application of the polaron approach by investigating the QD excited state population both as a function 
of pulse area $\Theta=\Omega\Delta\tau$ and within the time domain. 
In Fig.~{\ref{PolPlots}} we plot dynamics calculated using the 
polaron theory (dotted, blue curves), and the previous weak-coupling theory 
(dashed, red curves) for comparison, 
where we drive the QD at its polaron-shifted transition frequency. 
From the first row, corresponding to a low temperature regime in which $T=4~\mathrm{K}$, we see that both theories predict almost identical dynamics 
over a wide range of pulse areas (or equivalently driving strengths)~\cite{mccutcheon10_2}. 
To see why this is the case, for $\delta'=0$ the polaron 
Bloch equations can be solved for the QD population difference giving 
\begin{align}
\alpha_z(t)=\e^{-\gamma_{\mathrm{p}}t/2}
\Big[\cos\left(\frac{\xi_{\mathrm{p}} t}{2}\right)-\frac{\Gamma_{\mathrm{p1}}}{\xi_{\mathrm{p}}}\sin\left(\frac{\xi_{\mathrm{p}} t}{2}\right)\Big],
\label{alphazpol}
\end{align}
with damping rate $\gamma_{\rm p}=\Gamma_{\mathrm{p1}}+2\Gamma_{\mathrm{p3}}$ and oscillation frequency 
$\xi_{\mathrm{p}}=\sqrt{4\Omega_\pol(\Omega_\pol+\lambda_{\mathrm{p1}})-\Gamma_{\mathrm{p1}}^2}$. 
Since we are interested in the low temperature limit in which the single-phonon term should dominate, we now consider the polaron rate $\gamma_\pol$ 
to lowest order in the system--environment coupling strength $\alpha$. At this order the polaron 
correlation functions become $C_{xx}(\tau)\approx0$ and 
$C_{yy}\approx\av{B}^2 \phi(\tau)$, which leads to $\Gamma_{\mathrm{p3}}\approx0$ and thus  
$\gamma_{\rm p}\approx\Gamma_{\mathrm{p1}}$, which can be found analytically:
\beq
\gamma_{\rm p}\approx \frac{\pi}{2} J_{\rm ph}(\Omega_\pol)\coth(\beta\Omega_\pol/2).
\eeq
Additionally, in the limit $\Omega_\pol\ll k_B T,\w_c$ we may further simplify the rate to $\gamma_{\rm p}\approx \pi \alpha k_B T \Omega_\pol^2$. 
This demonstrates that in the weak-coupling limit, the damping rate in the polaron theory is the same as 
that in the weak-coupling theory, though evaluated at the renormalised Rabi frequency $\Omega_\pol=\Omega\av{B}$ 
rather than the original $\Omega$. Thus, both theories predict similar dynamics at low temperatures. 
The correspondence breaks down at higher temperatures, however, as Rabi frequency renormalisation (and multiphonon effects more generally) become significant. 
For example, in the middle row of Fig.~{\ref{PolPlots}}, for which $T=75~\mathrm{K}$, we see that while 
the weak-coupling theory predicts unphysical behaviour, 
the polaron master equation predictions remain physical.

Turning to the lower row, 
which corresponds to a temperature of 
$T=50~\mathrm{K}$ and for 
which we explore pulse areas as high as 
$\Theta=20\pi$, 
we see another way in which the polaron and weak-coupling theories can differ.
At small pulse areas, although differences are present, 
the polaron and weak-coupling theories predict similar qualitative behaviour. At larger 
pulse areas, however, the two theories begin to differ drastically in their predictions. For example, at $\Theta=15\pi$, 
corresponding to a driving strength of 
$\Omega=15 \pi/(10~\mathrm{ps})\approx 4.7~\mathrm{ps}^{-1}$ shown on the right, 
we see that the weak-coupling theory becomes undamped while the polaron theory 
is still strongly damped. This in fact demonstrates a failure of the polaron approach. 
For very large driving strengths exceeding the cut-off frequency of the phonon 
bath (here $\w_c=2.2~\mathrm{ps}^{-1}$), the phonon modes cannot follow the state 
of the QD and the exciton and phonons begin to decouple~\footnote{Though the relevant driving strengths are much larger than those that can presently be achieved under continuous-wave excitation conditions, peak Rabi frequencies under pulsed excitation can be significant in comparison to the phonon cut-off~\cite{ramsay10_2}.}. 
Though this effect is yet to be experimentally observed, it has also been predicted 
using powerful numerically exact path integral methods~\cite{vagov2007,mccutcheon11_2}. 
In this regime, owing to the decoupling effect, 
the weak-coupling theory then begins to become appropriate once again (as long as non-Markovian corrections are minor). The polaron 
theory, however, cannot capture this decoupling, since it assumes all modes can 
adiabatically follow the QD state. As such, as the driving strength is increased, it simply gives 
rise to larger and larger damping. We shall discuss these points further in the following section, as well as develop a refined version of the polaron master equation applicable also in the strong driving regime.
 

\section{QD Dynamics - Variational Polaron Theory}
\label{VariationalTheory}

In many parameter regimes the polaron master equation predicts dynamics more accurate than those 
of the weak-coupling theory. However, as we have just seen, for large driving strengths the polaron approach becomes invalid. This can be understood on a number of levels. For example, in Section~{\ref{PolaronTheory}} we saw how in the zero driving limit the 
polaron transformation exactly diagonalises the complete Hamiltonian. This inspired us to apply the transformation 
outside the zero driving limit, and to treat terms proportional to the Rabi frequency as a perturbation. In this case, 
it should come as little surprise that such a theory relies on the driving strength being weak. To give a more physical picture, 
the polaron transformation can be thought of as moving into a basis in which the environment oscillators are displaced 
in response to the QD state. If we take this displaced basis as a representation for our unperturbed Hamiltonian, 
then on an intuitive level we might expect this to be valid only for environment oscillators that are fast enough (i.e.~of sufficiently high frequency) to 
respond rapidly to the state of the QD. For large driving strengths, it seems likely that some (if not most) environment oscillators 
will be too sluggish to follow the QD state. This reasoning motivates us to explore a variational extension to the polaron theory~\cite{silbey84,harris85,silbey89,mccutcheon11_2}, in 
which the displacement we apply to each environment oscillator is chosen by some criterion that depends on the mode 
frequency, the driving strength, and the other system-environment parameters. 

\subsection{Variational polaron transformation}

To see how we determine the variational transformation, 
let us consider a 
generalisation of the polaron transformation 
introduced in Section~{\ref{PolaronTheory}}, which again acts on the full 
QD Hamiltonian in the rotating frame given in Eq.~({\ref{HTotal}}). We have 
\beq
H_\Var = \e^V H \e^{-V},
\label{HV0}
\eeq
where $V=\ketbra{X}{X}\sum_\kv f_\kv (b_\kv^{\dagger}-b_\kv)/\w_\kv$, and the set $\{f_\kv\}$ will be 
referred to as the variational parameters. With reference to 
Eq.~({\ref{HP0}}) we can see that for $f_\kv = g_\kv$ this is equivalent to the polaron transformation. 
The idea behind the variational theory, however, is to chose the $f_\kv$ 
in such a 
way that our perturbative master equation remains as accurate as possible, given the restricted form of transformation 
we consider. Applying the general transformation in Eq.~({\ref{HV0}}) we find~\cite{mccutcheon11_2}
\beq
H_\Var = H_{\mathrm{SV}}+H_{\mathrm{IV}}+H_{\mathrm{E}},
\eeq
where the system Hamiltonian is now given by
\beq
H_{\mathrm{SV}}=\delta_{\mathrm{v}} \ketbra{X}{X}+\frac{\Omega_{\mathrm{v}}}{2}\sigma_x,
\label{HSV}
\eeq
with $\delta_{\mathrm{v}}=\delta+R$ and $R=\sum_\kv f_\kv (f_\kv - 2 g_\kv)/\w_\kv$. In analogy with the previous 
polaron theory, a renormalised Rabi frequency is defined, given by $\Omega_{\mathrm{v}} = \Omega \av{\mathcal{B}}$ 
with $\av{\mathcal{B}}=\mathrm{Tr}_\Env[\mathcal{B}_{\pm} \rho_\Env ]$. Now the bath operators are functions 
of the variational parameters, $\mathcal{B}_{\pm}=\prod_\kv D_\kv(\pm f_\kv/\w_\kv)$, yielding 
the renormalisation factor 
\beq
\av{\mathcal{B}} = \exp\left[-\frac{1}{2}\sum_\kv \frac{f_\kv^2}{\w_\kv^2}\coth(\beta\w_\kv/2)\right],
\eeq
for a thermal state of the environment. 
The interaction Hamiltonian contains two terms, one polaron-like 
contribution and one weak-coupling-like contribution. Specifically,
\beq
H_{\mathrm{IV}}=\frac{\Omega}{2}\Big(\sigma_x \mathcal{B}_x + \sigma_y \mathcal{B}_y\Big)+\ketbra{X}{X}\mathcal{B}_z,
\label{HIV}
\eeq
where 
\begin{align}
\mathcal{B}_x&=\frac{1}{2}\big(\mathcal{B}_++\mathcal{B}_--2\av{\mathcal{B}}),\nonumber\\
\mathcal{B}_y&=\frac{1}{2i}\big(\mathcal{B}_--\mathcal{B}_+),
\label{BxByVar}
\end{align}
and 
\beq
\mathcal{B}_z=\sum_\kv (g_\kv - f_\kv)(b_\kv^{\dagger}+b_\kv).
\label{BzVar}
\eeq
Note that in general the bath operators in the variational theory, which we label $\mathcal{B}_x$, $\mathcal{B}_y$ and $\mathcal{B}_z$, 
are not equal to those encountered in the weak-coupling or polaron theories due to additional factors or terms involving the 
variational parameters. In the following we use calligraphic notation for bath operators and correlation functions in 
the variational frame.

\subsection{Free energy minimisation}

Before we go on to derive our variational master equation, we must first determine the 
free parameters $\{f_\kv\}$. We choose them such that they minimise the 
free energy associated with the variationally transformed Hamiltonian $H_\Var$. 
At zero temperature this corresponds to minimising the ground-state energy, as usual. 
At finite temperature, as the free energy is minimised in equilibrium,  
the variational transformation then attempts 
to find the best possible diagonalisation of the 
complete Hamiltonian, given the restricted form of this unitary~\cite{silbey84,harris85,silbey89,mccutcheon11_2}.

To minimise the free energy, we compute the Feynman-Bogoliubov upper bound given by 
\beq
A_{\mathrm{B}}=-\frac{1}{\beta} \mathrm{ln}(\mathrm{Tr}\{\e^{-\beta H_{\mathrm{0V}}} \})+\av{H_{\mathrm{IV}}}_{H_{\mathrm{0V}}}
+\mathcal{O}\av{H_{\mathrm{IV}}^2}_{H_{\mathrm{0V}}},
\label{ABDefinition}
\eeq
where $H_{\mathrm{0V}}=H_{\mathrm{SV}}+H_\Env$. The upper bound satisfies $A_{\mathrm{B}}\geq A$, with $A$ being 
the true free energy. We have constructed $H_{\mathrm{IV}}$ such that the second term in Eq.~({\ref{ABDefinition}}) 
is zero. Neglecting terms higher order in $H_{\mathrm{IV}}$ we find the approximate free energy bound
\beq
A_B\approx \frac{1}{2}(\delta_{\mathrm{v}}+\eta_{\mathrm{v}})-\frac{1}{\beta}\mathrm{ln}\big(1+\e^{\beta\eta_{\mathrm{v}}}\big),
\label{AB1}
\eeq
where $\eta_{\mathrm{v}}=\sqrt{\delta_{\mathrm{v}}^2+\Omega_{\mathrm{v}}^2}$, and we have neglected the contribution coming from the 
environment Hamiltonian, since it does not depend on the variational parameters. Differentiating Eq.~({\ref{AB1}}) 
with respect to $f_\kv$ we find 
\beq
\pder{A_B}{f_\kv}=\frac{1}{2}\pder{R}{f_\kv}-\frac{1}{2}\tanh(\beta\eta_{\mathrm{v}}/2)\pder{\eta_{\mathrm{v}}}{f_\kv},
\eeq
and solving the minimisation condition $\pder{A_B}{f_\kv}=0$ we obtain 
\beq
f_\kv =
\frac{g_\kv\big[1-\frac{\delta_{\mathrm{v}}}{\eta_{\mathrm{v}}}\tanh(\beta\eta_{\mathrm{v}}/2)\big]}
{1-\frac{\delta_{\mathrm{v}}}{\eta_{\mathrm{v}}}\tanh(\beta\eta_{\mathrm{v}}/2)\big[1-\frac{\Omega_{\mathrm{v}}^2}{2\delta_{\mathrm{v}}\w_\kv}\coth(\beta\w_\kv/2)\big]}.
\label{fMinimised}
\eeq
We emphasise that the variational parameters differ for each wavevector $\kv$. In particular, for those wavevectors $\kv$ having corresponding 
frequencies satisfying $\Omega_{\mathrm{v}}/\w_\kv \ll 1$, the 
minimisation condition approximates to $f_\kv \to g_\kv$. Thus, for these modes, the full polaron transformation 
should be applied. In contrast, for $\Omega_{\mathrm{v}}/\w_\kv \gg 1$ we find $f_\kv \to 0$, and no transformation 
is applied. This confirms our earlier intuition that for sluggish modes, for which $\Omega_{\mathrm{v}}/\w_\kv \gg 1$, 
the polaron transformation is not appropriate, since these modes cannot follow the state of the QD.

It is worth noting that the parameters 
$\delta_{\mathrm{v}}$, $\eta_{\mathrm{v}}$ and $\Omega_{\mathrm{v}}$ 
appearing in Eq.~({\ref{fMinimised}}) depend on 
the variational parameters $\{f_\kv\}$ themselves. As such, the renormalised quantities 
must typically be solved for self-consistently. In order to do so, we write 
$f_\kv=g_\kv F(\w_\kv)$ from which we can write $\delta_{\mathrm{v}} = \delta+R$ with 
\beq
R=\int_{0}^{\infty}\mathrm{d}\w \frac{J_{\rm ph}(\w) F(\w)}{\w}[F(\w)-2],
\eeq
and $\Omega_{\mathrm{v}} = \av{\mathcal{B}}\Omega$ with 
\beq\label{mathcalB}
\av{\mathcal{B}}=\exp\left[-\frac{1}{2}\int_0^{\infty}\mathrm{d}\w \frac{J_{\rm ph}(\w) F(\w)^2}{\w^{2}} \coth(\beta\w/2)\right],
\eeq
which we numerically solve simultaneously.

\subsection{Variational master equation}

To derive a master equation in the variational frame, we must first find the relevant correlation 
functions. From the interaction Hamiltonian, Eq.~({\ref{HIV}}), we see 
that within the variational representation there are two distinct contributions, one which resembles that 
in polaron theory, and another whose form is the same as in the weak-coupling approach. Accordingly,  
we expect the resulting master equation to have three contributions; terms which look similar to 
those found at weak-coupling, polaron-like terms, and cross terms arising 
from products of the two different types of bath operator. Indeed, 
this reflects the general nature of the variational method. In the appropriate limits it is expected to reduce 
to either the weak-coupling or the polaron theories, though in general both contributions will be present. 
The cross contributions are important when interpolating between these two cases. 

We label the variational frame interaction Hamiltonian system operators as $A_x=(\Omega/2)\sigma_x$, 
$A_y = (\Omega/2)\sigma_y$ and $A_z=\ketbra{X}{X}$. We must calculate the correlation 
functions $\mathcal{C}_{ij}(\tau)=\mathrm{Tr}_{\rm E}[\tilde{\mathcal{B}}_i(\tau) \mathcal{B}_j \rho_{\rm E}]$ for 
$i,j=\{x,y,z\}$, with bath operators given in Eqs.~({\ref{BxByVar}}) and ({\ref{BzVar}}). 
The correlation functions $\mathcal{C}_{xx}(\tau)$ and $\mathcal{C}_{yy}(\tau)$ are of precisely the same form as those 
encountered in polaron theory. In exact analogy with Eqs.~({\ref{Cxx}}) and ({\ref{Cyy}}) we find 
\begin{align}
\mathcal{C}_{xx}(\tau)&=\frac{\av{\mathcal{B}}^2}{2}\Big(\e^{\varphi(\tau)}+\e^{-\varphi(\tau)}-2\Big),\label{cxxvar}\\
\mathcal{C}_{yy}(\tau)&=\frac{\av{\mathcal{B}}^2}{2}\Big(\e^{\varphi(\tau)}-\e^{-\varphi(\tau)}\Big),\label{cyyvar}
\end{align}
and $\mathcal{C}_{xy}(\tau)=\mathcal{C}_{yx}(\tau)=0$, where the phonon propagator is now given by 
\beq
\varphi(\tau)=\int_0^{\infty}\!\!\!\mathrm{d}\w \frac{J_{\rm ph}(\w)}{\w^2}F(\w)^2\Big(\!\cos\w\tau\coth(\beta\w/2)-i\sin\w\tau\!\Big),
\eeq
and is thus dependent on the variational optimisation through $F(\omega)$.
Similarly, the weak-coupling-like correlation function is found to be
\begin{align}\label{czzvar}
\mathcal{C}_{zz}(\tau)=&\int_0^{\infty}\mathrm{d}\w J_{\rm ph}(\w) [1\!-\!F(\w)]^2\nonumber\\
&\times\Big(\!\cos\w\tau\coth(\beta\w/2)-i\sin\w\tau\!\Big).
\end{align}
The only correlation functions requiring additional effort arise from the cross terms. 
With the help of Appendix~{\ref{CorrelationFunctions}} [see Eq.~(\ref{Cyzsinglemode})] we find 
\begin{align}\label{cyzvar}
\mathcal{C}_{yz}(\tau)=-\av{\mathcal{B}}&\int_0^{\infty}\mathrm{d}\w \frac{J_{\rm ph}(\w)}{\w}F(\w)[1-F(\w)]\nonumber\\
&\times\Big(\! i \cos\w\tau+\sin\w\tau\coth(\beta\w/2)\!\Big),
\end{align}
while $\mathcal{C}_{zy}(\tau)=-\mathcal{C}_{yz}(\tau)$ and $\mathcal{C}_{zx}(\tau)=\mathcal{C}_{xz}(\tau)=0$.

Having found the relevant bath correlation functions we can now use Eq.~({\ref{rhodotD2}}) to write 
down the master equation in the variational frame. As in Section~\ref{PolaronTheory}, we must once again take care 
to remember that the Bloch equations we derive from the variational master equation contain expectation values in the 
variational frame, which for the coherences are related to those in the original frame by a factor of $\av{\mathcal{B}}$.
We derive Bloch equations of the form $\dot{\bm \alpha}_{\rm V}=M_{\rm V}\cdot{\bm \alpha}_{\rm V}+{\bm b}_{\rm V}$, with
\begingroup
\renewcommand*{\arraystretch}{2}
\begin{widetext}
\beq
M_V = -
\left( \begin{array}{ccc} 
\frac{\Omega_{\mathrm{v}}^2}{\eta_{\mathrm{v}}^2}\Gamma_{\mathrm{w1}}\!+\!\frac{\delta_{\mathrm{v}}^2}{\eta_{\mathrm{v}}^2}\Gamma_{\mathrm{w2}}\!+\!\Gamma_{\mathrm{p1}}\!+\!\frac{\Omega_{\mathrm{v}}}{\eta_{\mathrm{v}}}\Gamma_{\mathrm{v1}} & 
\delta_{\mathrm{v}} \!+\! \lambda_{\mathrm{w2}}\!+\!\frac{\delta_{\mathrm{v}}}{\eta_{\mathrm{v}}}\lambda_{\mathrm{p1}}\!-\!\frac{\delta_{\mathrm{v}} \Omega_{\mathrm{v}}}{\eta_{\mathrm{v}}^2}(\lambda_{\mathrm{v1}}\!-\!\lambda_{\mathrm{v2}}) & 
\frac{\delta_{\mathrm{v}}\Omega_{\mathrm{v}}}{\eta_{\mathrm{v}}^2}(\Gamma_{\mathrm{w1}}\!-\!\Gamma_{\mathrm{w1}})\!+\!\frac{\delta_{\mathrm{v}}}{2\eta_{\mathrm{v}}}\Gamma_{\mathrm{v1}}\!+\!\Gamma_{\mathrm{v2}}\\
-\delta_{\mathrm{v}}\!-\!\lambda_{\mathrm{w2}}\!-\!\frac{\delta_{\mathrm{v}}}{\eta_{\mathrm{v}}}\lambda_{\mathrm{p2}} & 
\frac{\Omega_{\mathrm{v}}^2}{\eta_{\mathrm{v}}^2}(\Gamma_{\mathrm{w1}}\!+\!\Gamma_{\mathrm{p3}})\!+\!\frac{\delta_{\mathrm{v}}^2}{\eta_{\mathrm{v}}^2}(\Gamma_{\mathrm{w2}}\!+\!\Gamma_{\mathrm{p2}})\!-\!\frac{\Omega_{\mathrm{v}}}{2\eta_{\mathrm{v}}}\Gamma_{\mathrm{v1}} & 
\Omega_{\mathrm{v}}\!+\!\frac{\Omega_{\mathrm{v}}}{\eta_{\mathrm{v}}}\lambda_{\mathrm{w1}}\!-\!\lambda_{\mathrm{v1}} \\
-\frac{\delta_{\mathrm{v}}\Omega_{\mathrm{v}}}{\eta_{\mathrm{v}}}(\Gamma_{\mathrm{p3}}\!-\!\Gamma_{\mathrm{p2}})\!-\!\Gamma_{\mathrm{v2}} & 
-\Omega_{\mathrm{v}}\!-\!\frac{\Omega_{\mathrm{v}}}{\eta_{\mathrm{v}}}\lambda_{\mathrm{p1}}\!+\!\frac{\Omega_{\mathrm{v}}^2}{\eta_{\mathrm{v}}^2}\lambda_{\mathrm{v1}}\!+\!\frac{\delta_{\mathrm{v}}^2}{\eta_{\mathrm{v}}^2}\lambda_{\mathrm{v2}} & 
 \Gamma_{\mathrm{p1}}\!+\!\frac{\delta_{\mathrm{v}}^2}{\eta_{\mathrm{v}}^2}\Gamma_{\mathrm{p2}}\!+\!\frac{\Omega_{\mathrm{v}}^2}{\eta_{\mathrm{v}}^2}\Gamma_{\mathrm{p3}}\!+\!\frac{\Omega_{\mathrm{v}}}{2\eta_{\mathrm{v}}}\Gamma_{\mathrm{v2}}
\end{array}
\right),
\eeq
\end{widetext}
\endgroup
\noindent being significantly more complicated than in either the weak-coupling or polaron cases. Here, the quantities $\Gamma_{\mathrm{w1}}$ etc.~with a `w' subscript are defined 
as in the weak-coupling theory of Section~{\ref{WeakCoupling}}, but with the true weak-coupling correlation function replaced by its variational theory counterpart.
Explicitly, 
these terms are defined as in Eqs.~({\ref{Gammaw1}}) to ({\ref{zetaw1}}), though with the correlation function replaced by Eq.~({\ref{czzvar}}). 
Similarly, the quantities with a `p' subscript are defined in Eqs.~({\ref{Gammap1}}) to ({\ref{zetap1}}), though calculated using the correlation functions in Eqs.~({\ref{cxxvar}}) and (\ref{cyyvar}). 
The remaining terms are unique to the variational theory, and are given by
\begin{align}
\Gamma_{\mathrm{v1}}&=\Omega(S_{zy}(\eta_{\mathrm{v}})-S_{zy}(-\eta_{\mathrm{v}})),\\
\Gamma_{\mathrm{v2}}&=\Omega S_{zy}(0),\\
\lambda_{\mathrm{v1}}&=\sfrac{\Omega}{4}(\gamma_{zy}(\eta_{\mathrm{v}})+\gamma_{zy}(-\eta_{\mathrm{v}})),\\
\lambda_{\mathrm{v2}}&=\sfrac{\Omega}{2}\gamma_{zy}(0).
\end{align}
The 
inhomogeneous terms are 
\beq
\vec{b}_\Var \!= \!\!\left( \begin{array}{c}
\!\!-\frac{\Omega_{\mathrm{v}}}{\eta_{\mathrm{v}}} \big[\kappa_{\mathrm{w1}}+\kappa_{\mathrm{p1}}\big]-\frac{\Omega_{\mathrm{v}}^2+\eta_{\mathrm{v}}^2}{2\eta_{\mathrm{v}}^2}\kappa_{\mathrm{v1}} - \frac{\delta_{\mathrm{v}}^2}{\eta_{\mathrm{v}}^2}\Gamma_{\mathrm{v2}}\!\\
\!\!-\frac{\Omega_{\mathrm{v}}\delta_{\mathrm{v}}}{\eta_{\mathrm{v}}^2}\big[\zeta_{\mathrm{w1}}-\zeta_{\mathrm{p1}}-\lambda_{\mathrm{w2}}+\lambda_{\mathrm{p3}}\big] +\frac{\delta_{\mathrm{v}}}{2\eta_{\mathrm{v}}}\zeta_{\mathrm{v}}\!\\
\!\!-\frac{\delta_{\mathrm{v}}}{\eta_{\mathrm{v}}}\big[\kappa_{\mathrm{p1}}+\kappa_{\mathrm{p2}}\big]-\frac{\delta_{\mathrm{v}}\Omega_{\mathrm{v}}}{2\eta_{\mathrm{v}}^2}\big[\kappa_{\mathrm{v1}}-2\Gamma_{\mathrm{v2}}\big]\!
 \end{array}
\right),
\eeq
%
with coefficients containing a `w' or `p' subscript defined as above, i.e.~the corresponding weak-coupling or polaron expression with the bath correlation function replaced by the appropriate variational form, and 
\begin{align}
\kappa_{\mathrm{v1}}=\Omega(S_{zy}(\eta_{\mathrm{v}})+S_{zy}(-\eta_{\mathrm{v}})),\\
\zeta_{\mathrm{v1}}=\sfrac{\Omega}{2}(\gamma_{zy}(\eta_{\mathrm{v}})-\gamma_{zy}(-\eta_{\mathrm{v}})).
\end{align}
In the limit that $F(\omega)\rightarrow1$ all terms with a `w' or `v' subscript disappear, and the Bloch equations reduce to the polaron form as expected. In the opposite limit, $F(\omega)\rightarrow0$ all `p' and `v' terms vanish and we recover the weak-coupling Bloch equations.

\begin{figure}
\begin{center}
\includegraphics[width=0.49\textwidth]{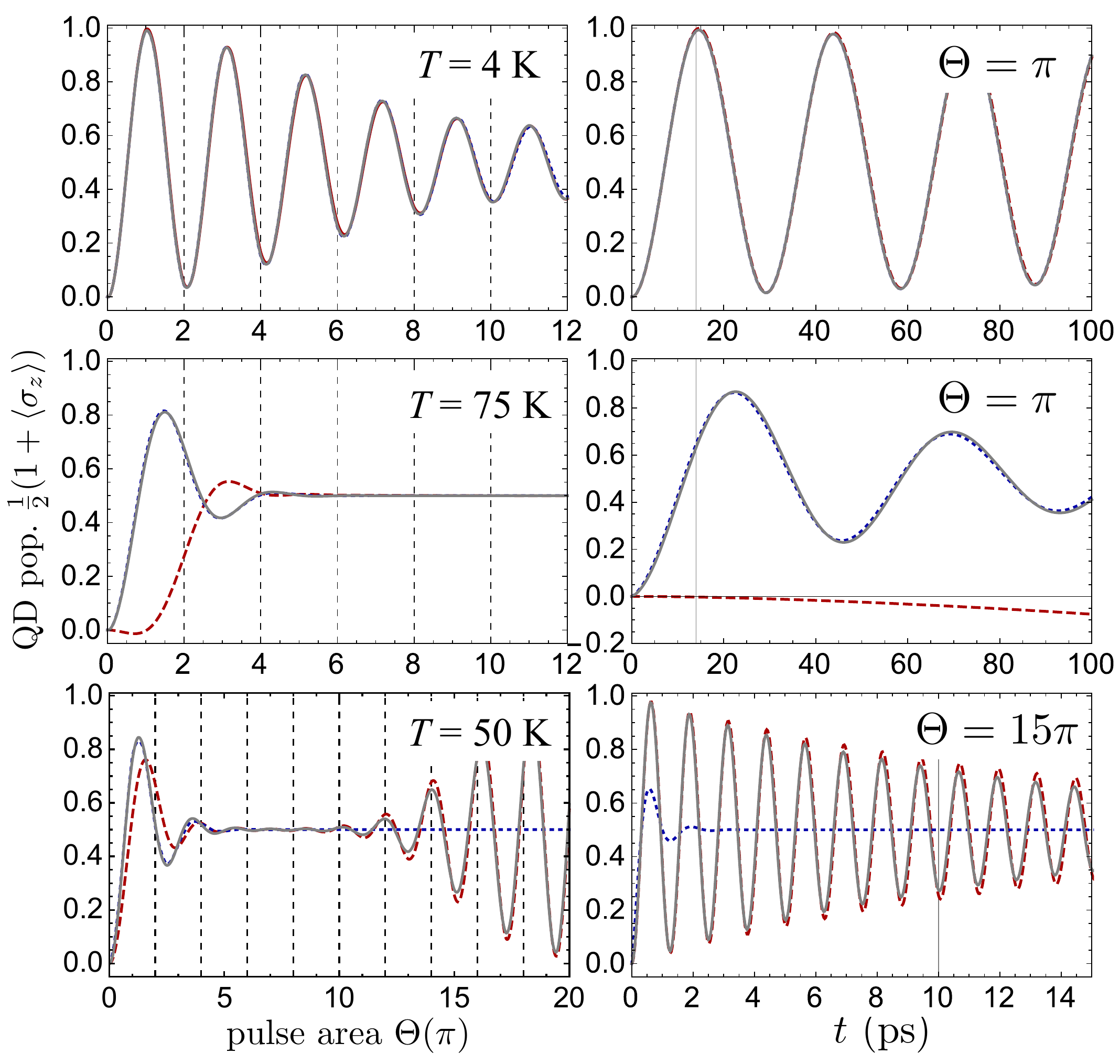}
\caption{QD excited state population as a function of pulse area (left) and 
time (right), calculated using the variational theory (solid, grey curves), polaron theory (dotted, blue curves) and the 
weak-coupling theory (dashed, red curves). For low temperature all three theories predict similar behaviour, 
while for $T=75~\mathrm{K}$ and moderate driving strengths the variational theory 
mimics the polaron theory in order to minimise perturbative 
terms in the transformed Hamiltonian. For very large driving strengths the variational and 
weak-coupling theories approximately coincide as the exciton and phonons decouple.}
\label{VarPlots}
\end{center}
\end{figure}

In Fig.~{\ref{VarPlots}} we compare the dynamics calculated using the variational Bloch equations (solid, grey curves), 
with those of the polaron (dotted, blue curves) and weak-coupling (dashed, red curves) techniques. As in 
the previous cases, we plot the QD population as a function of pulse area on the left, and 
in the time domain on the right. At low temperatures and pulse areas of up to at least $\Theta=12\pi$ 
(corresponding to driving strengths of $\Omega=\Theta/\Delta\tau\approx 2.7~\mathrm{ps}^{-1}$), 
it can be seen that all three theories predict equivalent dynamics. 
For these parameters 
the variational transformation is approximately equal to the full polaron transformation, which as we have seen 
gives behaviour close to the weak-coupling theory for low enough temperatures. 
In the middle row, where $T=75~\mathrm{K}$, we again find that the variational and polaron theories match, 
but here the weak-coupling Bloch equations deviate markedly, 
predicting unphysical behaviour as previously outlined. The variational transformation in this high temperature 
regime thus corresponds closely to the full polaron transformation (except for very low frequency modes), 
since such a choice acts to minimise the perturbative terms in the Hamiltonian. These terms are, however, now large (and hence non-perturbative) if no transformation is applied, which is the case in the weak-coupling theory.

The real versatility of the variational approach can be appreciated when considering the lowest 
row in Fig.~\ref{VarPlots}, where Rabi frequencies of up to $\Omega=20 \pi/(10~\mathrm{ps})\approx 6.3~\mathrm{ps}^{-1}$ are explored at $T=50$~K. 
In contrast to the incorrect predictions of the polaron approach, at very strong driving the variational theory is able to capture the exciton-phonon decoupling effect seen also in the weak-coupling dynamics.
In this regime, the full polaron transformation  
becomes inappropriate as it relies on displacing the phonons such that they adiabatically follow the QD state. However, for strong driving, the majority of important phonon modes become sluggish with respect to the QD Rabi frequency and they should therefore be displaced only by a small amount, or even not at all. The variational 
transformation then naturally begins to shift towards the identity (i.e.~no transformation), and instead we end up performing 
perturbation in the original system-environment coupling strength. Thus, we see that by attempting to minimise the interaction terms, the variational formalism is able to interpolate between regimes in which the polaron representation is advantageous, and those in which a weak-coupling approach is more favourable.


\section{Phonon effects in quantum dot photon emission}
\label{Emission}

Our considerations so far have 
focussed on the influence of phonon interactions on the QD exciton population and coherence dynamics. Direct access to the excitonic population can be gained, for example, through photocurrent measurements~\cite{zrenner02,ramsay10}. Furthermore, QD populations and coherences can conveniently be calculated directly from the master equations that we have derived. However, it is often preferable to probe QD systems via measurements on their emitted photons, the characterisation of which is also vitally important in the development of new QD photonic technologies. Such measurements also provide access to information beyond simply elements of the QD reduced density matrix, such as multi-time correlation functions and emission spectra. More generally, it is an interesting and important problem to understand how standard quantum optics techniques and intuition must be modified to account for the solid-state nature of emitters such as semiconductor QDs. Hence, the focus of this section will be on developing a means to characterise the phonon influence in QD photon emission properties.

\subsection{Including both spontaneous emission and phonon interactions}

We have 
treated interactions between the QD excitonic degrees of freedom and the phonon and photon environments separately up to now. However, we must account for both phonon coupling and photon emission simultaneously if we are to characterise the effect of the former on the latter. Let us start from the Hamiltonian in the rotating frame
\begin{align}\label{Hphononphoton}
H = H_{\rm S} + H_{\rm E_{1}}+H_{\rm E_{2}}+H_{\rm I_{1}}+H_{\rm I_{2}},
\end{align}
where
\begin{align}\label{HSRWA}
H_{\rm S}=\delta\ket{X}\bra{X}+\frac{\Omega}{2}(\ket{0}\bra{X}+\ket{X}\bra{0}),
\end{align}
again describes a classically driven QD within the RWA [see Eq.~(\ref{hdrivenrwa})],
\begin{align}\label{Henvs}
H_{\rm E_{1}}&=\sum_{\bf k}\omega_{\bf k}b_{\bf k}^{\dagger}b_{\bf k},\\
H_{\rm E_{2}}&=\sum_{\bf q}\nu_{\bf q}a_{\bf q}^{\dagger}a_{\bf q},
\end{align}
are the free phonon and photon Hamiltonians, respectively, and 
\begin{align}\label{Hints}
H_{\rm I_{1}}&=|X\rangle\langle X|\sum_{\bf k}g_{\bf k}(b_{\bf k}^{\dagger}+b_{\bf k}),\\
H_{\rm I_{2}}&=\sum_{\bf q}u_{\bf q}(e^{-i\omega_l t}|0\rangle\langle X|a_{\bf q}^{\dagger}+e^{i\omega_l t}|X\rangle\langle 0|a_{\bf q}),
\end{align}
define the QD exciton-phonon and exciton-photon couplings, respectively.

We shall restrict ourselves to the variational treatment of the phonon interaction term, as it is the most general and both the weak-coupling and polaron methods can also be recovered from it in the appropriate limits. Applying the unitary transformation as in Section~\ref{VariationalTheory} to Eq.~(\ref{Hphononphoton}) we obtain
\begin{align}\label{Hphononphotonvar}
H_{\rm V} = H_{\rm SV} + H_{\rm E_{1}}+H_{\rm E_{2}}+H_{\rm IV_{1}}+H_{\rm IV_{2}}.
\end{align}
Here, $H_{\rm SV}$ is as given in Eq.~(\ref{HSV}), 
$H_{\rm IV_{1}}$ is the exciton-phonon interaction term in the transformed representation given in Eq.~(\ref{HIV}), and 
\begin{align}
H_{\rm IV_{2}}=\sum_{\bf q}u_{\bf q}(e^{-i\omega_l t}{\mathcal B}_-|0\rangle\langle X|a_{\bf q}^{\dagger}+e^{i\omega_l t}{\mathcal B}_+|X\rangle\langle 0|a_{\bf q}),
\end{align}
now contains QD, phonon, and photon operators and describes how the exciton-photon coupling becomes modified after the variational transformation.
Note that if we assume the full environmental state to be initially thermal then it is also separable, $\rho_{\rm E}(0)=\rho_{\rm E_{1}}(0)\rho_{\rm E_{2}}(0)$, with $\rho_{\rm E_{1}}(0)$ and $\rho_{\rm E_{2}}(0)$ describing phonon and photon bath thermal states, respectively.
Hence, $\langle{\mathcal B}\rangle$ can be defined exactly as in Eq.~(\ref{mathcalB}) and ${\rm Tr}_{\rm E}[H_{\rm IV_{1}}\rho_{\rm E}(0)]={\rm Tr}_{\rm E}[H_{\rm IV_{2}}\rho_{\rm E}(0)]=0$.

We now move into the interaction picture with respect to $H_{\rm 0V}=H_{\rm SV} + H_{\rm E_{1}}+H_{\rm E_{2}}$ such that
\begin{align}\label{HIVsum}
\tilde{H}_{\rm IV}(t)=\tilde{H}_{\rm IV_{1}}(t)+\tilde{H}_{\rm IV_{2}}(t)
\end{align}
where $\tilde{H}_{{\rm IV}_{j}}(t)=e^{iH_{\rm 0V}t}{H}_{{\rm IV}_{j}}e^{-iH_{\rm 0V}t}$ for $j=\{1,2\}$. From Eq.~(\ref{MEinteraction}) we know that our second-order master equation may be written as 
\begin{align}
\frac{\mathrm{d}}{\mathrm{d} t}\tilde{\rho}_{\rm SV}(t)=-&\int_0^{t}\mathrm{d}t_1
{\rm Tr}_{\rm E}[\tilde{H}_{\rm IV}(t),[\tilde{H}_{\rm IV}(t_1),\tilde{\rho}_{\rm SV}(t)\rho_{\rm E}(0)]].
\end{align} 
Inserting $\tilde{H}_{\rm IV}(t)=\tilde{H}_{\rm IV_{1}}(t)+\tilde{H}_{\rm IV_{2}}(t)$ we find that the master equation consists of four terms corresponding to the different possible combinations of the two interaction terms. However, as ${\rm Tr}_{\rm E}[H_{\rm IV_{1}}\rho_{\rm E}(0)]={\rm Tr}_{\rm E}[H_{\rm IV_{2}}\rho_{\rm E}(0)]=0$, those terms containing both $\tilde{H}_{\rm IV_{1}}(t)$ and $\tilde{H}_{\rm IV_{2}}(t)$ disappear. Thus, we obtain
\begin{align}\label{MEphononphoton}
\dot{\rho}_{\rm SV}(t)
=&-i[H_{\rm SV},{\rho}_{\rm SV}(t)]\nonumber\\
&-\int_0^{t}\mathrm{d}\tau{\rm Tr}_{\rm E_1}[{H}_{{\rm IV}_1},[\tilde{H}_{{\rm IV}_1}(-\tau),{\rho}_{\rm SV}(t)\rho_{\rm E_1}(0)]]\nonumber\\
&-\int_0^{t}\mathrm{d}\tau
{\rm Tr}_{\rm E}[{H}_{{\rm IV}_2},[\tilde{H}_{{\rm IV}_2}(-\tau),{\rho}_{\rm SV}(t)\rho_{\rm E}(0)]],
\end{align} 
where we have transformed back to the Schr\"odinger picture. 
Here, the first two terms are precisely those that we obtained in the previous variational master equation when considering only phonon interactions, and are thus unaffected by the extra coupling to the radiation field. The third term, responsible for photon emission and absorption processes due to the radiation field, appears at first sight to be modified by the phonon environment. However, we shall now see that for typical QD parameters this term should actually reduce to the usual form expected in the absence of phonon interactions. 

From Eq.~(\ref{MEphononphoton}), consider the last term on the right hand side,
\begin{align}\label{RHS2}
-\int_0^{t}\mathrm{d}\tau{\rm Tr}_{\rm E_1+E_2}[{H}_{{\rm IV}_2},[\tilde{H}_{{\rm IV}_2}(-\tau),{\rho}_{\rm SV}(t)\rho_{\rm E_1}(0)\rho_{\rm E_2}(0)]],
\end{align}
where we have
\begin{align}
\tilde{H}_{\rm IV_{2}}(t)=&\sum_{\bf q}u_{\bf q}\Big(e^{-i\omega_l t}{\mathcal B}_-(t)\sigma_-(t)a_{\bf q}^{\dagger}e^{i\nu_{\bf q}t}\nonumber\\
&+e^{i\omega_l t}{\mathcal B}_+(t)\sigma_+(t)a_{\bf q}e^{-i\nu_{\bf q}t}\Big)\nonumber\\
&=\sum_{i=1}^{2}\tilde{A}_i(t)\otimes\tilde{B}_i(t)\otimes\tilde{C}_i(t),
\end{align}
with $\tilde{A}_1(t)=\sigma_-(t)e^{-i\omega_l t}$, $\tilde{A}_2(t)=\tilde{A}_1^{\dagger}(t)$, $\tilde{B}_1(t)={\mathcal B}_{-}(t)$, $\tilde{B}_2(t)=\tilde{B_1}^{\dagger}(t)$, $\tilde{C_1}(t)=\sum_{\bf q}u_{\bf q}a_{\bf q}^{\dagger}e^{i\nu_{\bf q}t}$, and $\tilde{C_2}(t)=\tilde{C_1}^{\dagger}(t)$. We have also defined $\sigma_{-}(t)=e^{iH_{\rm 0V}t}\sigma_{-}e^{-iH_{\rm 0V}t}$, with $\sigma_-=|0\rangle\langle X|$, and ${\mathcal B}_{-}(t)=e^{iH_{\rm 0V}t}{\mathcal B}_{-}e^{-iH_{\rm 0V}t}$. Though this interaction Hamiltonian is more complicated than the form previously considered in Eq.~(\ref{intdecomp}), the product initial state of the environment means that we may still write Eq.~(\ref{RHS2}) in the standard form
\begin{align}\label{RHS2standard}
-\sum_{ij}\int_0^{t}\mathrm{d}\tau\Big(&C'_{ij}(\tau)[A_{i},\tilde{A}_{j}(-\tau){\rho}_{\rm SV}(t)]\nonumber\\
&+C'_{ji}(-\tau)[\rho_{\rm SV}(t)\tilde{A}_{j}(-\tau),A_{i}]\Big),
\end{align}
where the correlation functions are now defined as 
\begin{align}\label{photonphononcorr}
C'_{ij}(\tau)={\rm Tr}_{\rm E_1}[\tilde{B}_i(\tau)B_j\rho_{\rm E_1}(0)]{\rm Tr}_{\rm E_2}[\tilde{C}_i(\tau)C_j\rho_{\rm E_2}(0)].
\end{align}
If we consider the initial photon field state to be the multimode vacuum, which is an excellent approximation under standard experimental conditions, then (as in Section~\ref{sebackground}) the only non-zero correlation function is 
\begin{align}\label{photonphononcorrnonzero}
C'_{21}(\tau)&={\rm Tr}_{\rm E_1}[\tilde{B}_2(\tau)B_1\rho_{\rm E_1}(0)]{\rm Tr}_{\rm E_2}[\tilde{C}_2(\tau)C_1\rho_{\rm E_2}(0)]\nonumber\\
&={\mathcal C}_{+-}(\tau)\int_0^{\infty}\mathrm{d}\nu J_{\rm pt}(\nu)e^{-i\nu\tau},
\end{align}
where ${\mathcal C}_{+-}(\tau)={\rm Tr}_{\rm E_1}[{\mathcal B}_+(\tau){\mathcal B}_-\rho_{\rm E_1}(0)]$, we have taken the continuum limit of the photon field, and defined the photon environment spectral density $J_{\rm pt}(\nu)$, see Eqs.~(\ref{continuumlimit}) and~(\ref{specdensdefn}). For the free field case in which we are interested, we may approximate this spectral density to be flat around frequencies of interest and hence replace it simply by a constant $J_{\rm pt}(\nu)\approx\kappa$ (corresponding to a Markov approximation in the time domain). Additionally, we extend the lower limit of integration to $-\infty$ under the assumption that only frequencies close to the QD resonance are important. 
Hence, 
\begin{align}\label{photonphononcorrnonzero}
C'_{21}(\tau)&\approx{\mathcal C}_{+-}(\tau)\kappa\int_{-\infty}^{\infty}\mathrm{d}\nu e^{-i\nu\tau}\nonumber\\
&=2\pi\kappa\;{\mathcal C}_{+-}(\tau)\delta(\tau).
\end{align}
Inserting this correlation function into Eq.~(\ref{RHS2standard}) and performing the integration over $\tau$ we obtain the standard form of dissipator for spontaneous emission
\begin{align}
\gamma'\left(\sigma_-\rho_{\rm SV}(t)\sigma_+-(1/2)\{\sigma_+\sigma_-,\rho_{\rm SV}(t)\}\right),
\end{align}
with rate
\begin{align}
\gamma'=2\pi\kappa\;{\mathcal C}_{+-}(0).
\end{align}
However, since ${\mathcal C}_{+-}(0)=1$ (see Appendix~\ref{CorrelationFunctions}), we find that the emission rate is unaltered from that in the absence of the phonon environment, $\gamma'\rightarrow\gamma=2\pi\kappa$.

Thus, for a QD exciton coupled to both the phonon environment and the free vacuum electromagnetic field, we may separate the two processes in our second-order master equation, considering each to be independent of the other. This is valid regardless of whether we treat the phonons within the weak-coupling, polaron, or variational representations.

\subsection{The output field}

We are now in a position to explore the QD dynamics in the presence of both phonon and photon environments. However, in order to use the master equation formalism to probe the optical properties of the system, we still need to relate the field emitted by the QD to its internal degrees of freedom. Consider the electric field operator at the origin 
\begin{align}
{\bf E}(t)=\sum_{\bf q}{\bm \epsilon}_{\bf q}\sqrt{\frac{\omega_{\bf q}}{2\epsilon_0V}}\left[a_{\bf q}(t)+a_{\bf q}^{\dagger}(t)\right],
\end{align}
which we write as ${\bf E}(t)={\bf E}_+(t)+{\bf E}_-(t)$. Here,
\begin{align}
{\bf E}_+(t)=\sum_{\bf q}E^0_{\bf q}{\bm\epsilon}_{\bf q}a_{\bf q}(t),
\end{align}
${\bf E}_-(t)={\bf E}_+^{\dagger}(t)$, and we have defined the field amplitude $E^0_{\bf q}$. We would like to relate the field operators to the QD internal degrees of freedom that are tracked in our master equations. To do so, consider the Heisenberg equations of motion for the field operators generated from the full Hamiltonian given in Eq.~(\ref{Hphononphoton}),
\begin{align}
\frac{\mathrm{d}}{\mathrm{d} t}a_{\bf q}(t)=-i\nu_{\bf q}a_{\bf q}(t)-iu_{\bf q}e^{-i\omega_lt}\sigma_-(t),
\end{align}
which we may formally integrate to give
\begin{align}
a_{\bf q}(t)=e^{-i\nu_{\bf q}t}a_{\bf q}(0)-i\int_0^tdt'u_{\bf q}e^{-i\omega_lt'}\sigma_-(t')e^{i\nu_{\bf q}(t'-t)}.
\end{align}
We then find the positive frequency component of the emitted field to be
\begin{align}
{\bf E}_+(t)=&\sum_{\bf q}E^0_{\bf q}{\bm\epsilon}_{\bf q}e^{-i\nu_{\bf q}t}a_{\bf q}(0)\nonumber\\
&-i\sum_{\bf q}E^0_{\bf q}{\bm\epsilon}_{\bf q}\int_0^tdt' u_{\bf q}e^{-i\omega_lt'}\sigma_-(t')e^{i\nu_{\bf q}(t'-t)}.
\end{align}
Here, the first term is the free evolution of the field that would be obtained in the absence of the QD, which remains in the vacuum state. It does not, therefore, contribute to the field correlation functions in which we shall be interested and can thus be ignored from now on. 
As in the previous section, we take the continuum limit, assume that the coupling is approximately constant around the frequencies of interest, and extend the lower limit of integration over frequency to $-\infty$, such that
\begin{align}\label{Eplust}
E_+(t)&\approx
-i{E}^0\sqrt{\kappa}\int_0^tdt'\int_{-\infty}^{\infty}\mathrm{d}\nu e^{-i\omega_lt'}\sigma_-(t')e^{i\nu(t'-t)}\nonumber\\
&=-2i\pi{E}^0\sqrt{\kappa}\int_0^tdt'e^{-i\omega_lt'}\sigma_-(t')\delta(t-t')\nonumber\\
&=-i\pi{E}^0\sqrt{\kappa}e^{-i\omega_lt}\sigma_-(t).
\end{align}
Note that we are now neglecting the mode polarisation vectors, which give rise to geometric factors in the field correlation functions in which we shall be interested, but do not change their qualitative behaviour~\cite{carmichael,stecknotes}. As a final remark, it is worth stressing that we have established a relationship between the emitted field and the QD internal dynamics based on the original Hamiltonian of Eq.~(\ref{Hphononphoton}). Working in the variational picture by starting from Eq.~(\ref{Hphononphotonvar}), the resulting expressions would also contain factors of the phonon displacements ${\mathcal B}_{\pm}(t)$. This can in turn impact upon the short time behaviour of the field correlation functions, though we shall not consider such complications in the following.

\subsection{Emission spectra}

By relating the optical field emitted from the QD to its internal dynamical evolution, we may now use our master equation techniques to study how phonon interactions impact upon the QD optical emission characteristics. As an example, we shall consider the 
QD emission spectrum 
under resonant driving conditions (resonance fluorescence), 
and explore how this departs from well-known results obtained in the atomic case, for which the phonon environment is absent.

From the (optical) Wiener-Khinchin theorem we may write the QD emission intensity spectrum as~\cite{stecknotes} 
\begin{align}\label{emissionspectrum}
I(\omega)&=\frac{1}{2\pi}\int_{-\infty}^{\infty}\mathrm{d}\tau\langle E_-(t)E_+(t+\tau)\rangle e^{i\omega\tau}\nonumber\\
&\propto\frac{1}{2\pi}\int_{-\infty}^{\infty}\mathrm{d}\tau\langle \sigma_+(t)\sigma_-(t+\tau)\rangle e^{i(\omega-\omega_l)\tau},
\end{align}
where we haved used Eq.~(\ref{Eplust}) in the second line. Taking the long time limit, $t\rightarrow\infty$, we define the spectral component as the Fourier transform of the QD first order correlation function, $g^{(1)}(\tau)$, such that
\begin{align}\label{spectrumdefn}
S(\omega)=\frac{1}{2\pi}\int_{-\infty}^{\infty}\mathrm{d}\tau g^{(1)}(\tau) e^{i(\omega-\omega_l)\tau},
\end{align}
where 
\begin{align}
g^{(1)}(\tau)&={\rm lim}_{t\rightarrow\infty}\langle\sigma_+(t)\sigma_-(t+\tau)\rangle\nonumber\\
&={\rm lim}_{t\rightarrow\infty}{\rm Tr}[\sigma_+(t)\sigma_-(t+\tau)\rho(0)]\nonumber\\
&={\rm lim}_{t\rightarrow\infty}{\rm Tr}[\sigma_+\sigma_-(\tau)\rho(t)]\nonumber\\
&=\langle\sigma_+\sigma_-(\tau)\rangle_{\rm ss}.
\end{align}
We may calculate QD two-time correlation functions such as $g^{(1)}(\tau)$ using the quantum regression theorem,  
which tells us 
how to find them 
directly from the master equation~\cite{carmichael,stecknotes}. To see this, consider a correlation function of the form
\begin{align}
\langle S_1(t)S_2(t+\tau)\rangle&={\rm Tr}[S_1(t)S_2(t+\tau)\rho(0)]\nonumber\\
&= {\rm Tr_S}[S_2\Lambda(t,\tau)],
\end{align}
where we have used the cyclic invariance of the trace and defined an effective reduced density operator
\begin{align}\label{reducedeffective}
\Lambda(t,\tau)&={\rm Tr_E}[e^{-i{\mathcal H}\tau}\rho(t)S_1e^{i{\mathcal H}\tau}],
\end{align}
with ${\mathcal H}$ the system-environment Hamiltonian under consideration. 
Differentiating Eq.~(\ref{reducedeffective}) with respect to $\tau$, we find that we can derive a second-order 
master equation for $\Lambda(t,\tau)$ that has precisely the same form as the respective master equation for the reduced density operator $\rho_{\rm S}(t)$, provided that we assume the Born-Markov approximations hold such that $\rho(t)\approx\rho_{\rm S}(t)\rho_E(0)$; i.e.~the environment remains 
in its initial (usually equilibrium) state~\cite{McCutcheon2015,Goan2011}. Note that this does not necessarily impose a weak-coupling limitation if we are working in either the polaron or variational formalism. Thus, the same equations of motion that we use to propagate the reduced density operator may also be applied to find two-time correlation functions, now subject to the initial condition $\Lambda(t,0)=\rho_{\rm S}(t)S_1$. 
Similar arguments can also be used to find higher-order correlation functions. 

\begin{figure}[t!]
\begin{center}
\includegraphics[width=0.49\textwidth]{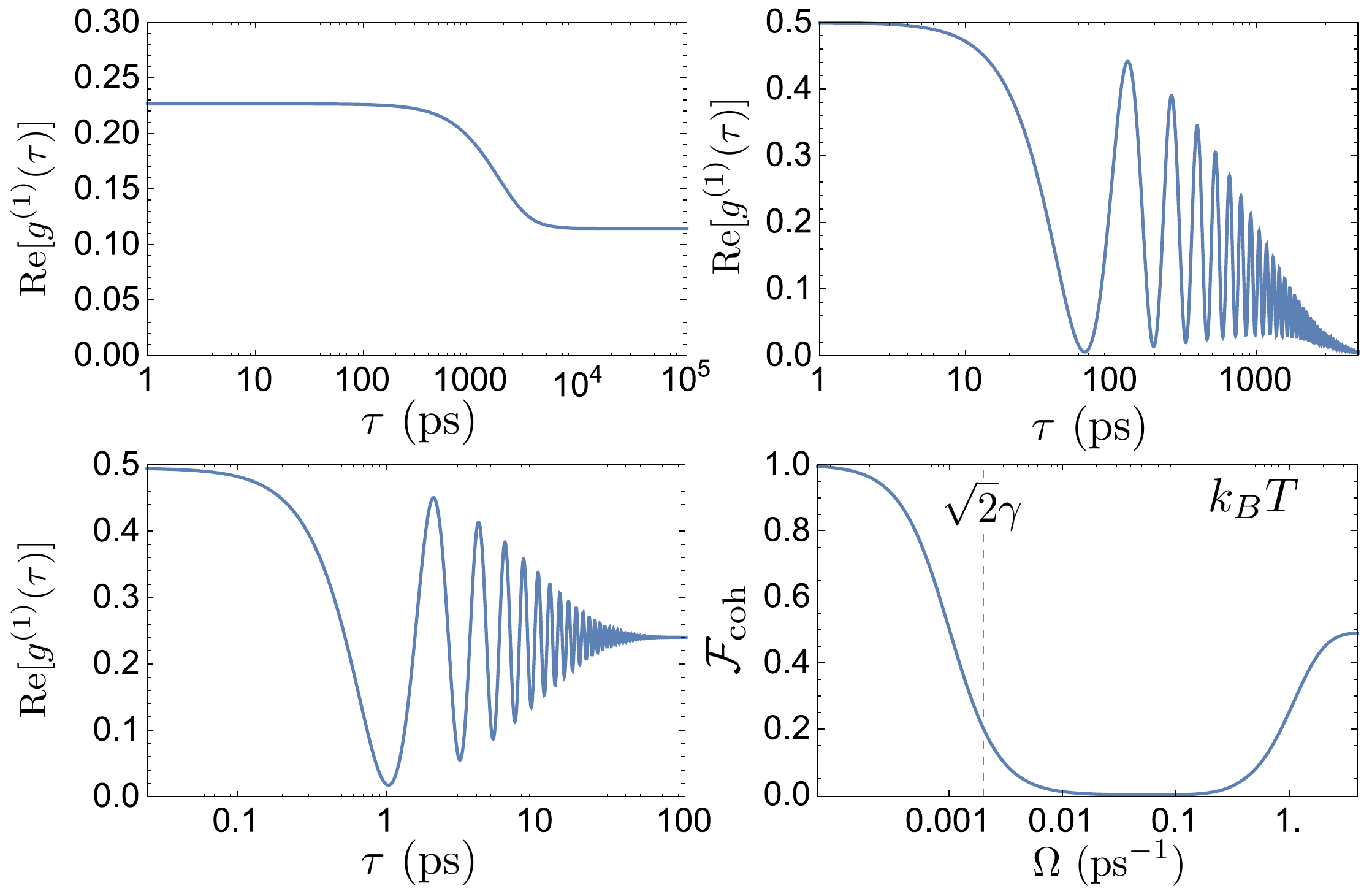}
\caption{Top and bottom left: Dynamical evolution of the QD first-order field correlation function, calculated from the variational master equation, for driving strengths $\Omega=0.001$~ps$^{-1}$ (top left), $\Omega=0.05$~ps$^{-1}$ (top right), and $\Omega=3.0$~ps$^{-1}$ (bottom left). 
Bottom right: Fraction of coherent emission as a function of driving strength. Notice the departure from textbook quantum optics with the reemergence of coherent scattering around $\Omega\sim k_BT$, which occurs only when phonon effects are treated rigorously~\cite{McCutcheon2013}. 
Other parameters used: $T=4~\mathrm{K}$, 
$\alpha=0.027~\mathrm{ps}^2$, $\w_c=2.2~\mathrm{ps}^{-1}$, $T_1=1/\gamma=700$~ps, 
and we drive the QD at the polaron-shifted resonance, $\delta=\int_0^{\infty}\mathrm{d}\w J_{\rm ph}(\w)/\w$.}
\label{g1}
\end{center}
\end{figure}

Returning to the particular case of QD emission, the correlation function is then given by 
\begin{align}
g^{(1)}(\tau)={\rm Tr_S}[\sigma_-\Lambda(\tau)]=\langle X| \Lambda(\tau)|0\rangle=\Lambda_{X0}(\tau),
\end{align} 
where $\Lambda(\tau)$ satisfies the master equation under consideration for initial condition $\Lambda(0)=\rho_{\rm S}(\infty)\sigma_+$, with $\rho_{\rm S}(\infty)={\rm lim}_{t\rightarrow\infty}[\rho_{\rm S}(t)]$ being the long-time (steady-state) system density operator. Notice that ${\rm Tr_S}[\Lambda(0)]={\rm Tr_S}[\sigma_+\rho_{\rm S}(\infty)]=\langle\sigma_+\rangle_{\rm ss}$, such that the effective reduced density operator is unnormalised. Hence, for large delay times $\tau$ the correlation function factorises as we might expect,
\begin{align}
{\rm lim}_{\tau\rightarrow\infty}[g^{(1)}(\tau)]=\langle\sigma_+\rangle_{\rm ss}\langle\sigma_-\rangle_{\rm ss}=|\rho_{0X}(\infty)|^2,
\end{align}   
and is equal to the magnitude of the off-diagonals of the steady-state QD reduced density matrix. A finite magnitude steady-state QD coherence thus implies a finite level of first-order coherent photon scattering. 

We can make this explicit in the spectrum by writing it as the sum of two terms,
\begin{align}
S(\omega)=S_{\rm coh}(\omega)+S_{\rm inc}(\omega),
\end{align}
with
\begin{align}
S_{\rm coh}(\omega)&=\frac{1}{2\pi}\int_{-\infty}^{\infty}\mathrm{d}\tau \langle\sigma_+\rangle_{\rm ss}\langle\sigma_-\rangle_{\rm ss} e^{i(\omega-\omega_l)\tau},\\
S_{\rm inc}(\omega)&=\frac{1}{2\pi}\int_{-\infty}^{\infty}\mathrm{d}\tau \langle{\sigma}_+'{\sigma}_-'(\tau)\rangle_{\rm ss}e^{i(\omega-\omega_l)\tau}.
\end{align}
Here we have defined 
\begin{align}
{\sigma}_+'(t)&=\sigma_+(t)-\langle\sigma_+\rangle_{\rm ss},\\
{\sigma}_-'(t)&=\sigma_-(t)-\langle\sigma_-\rangle_{\rm ss},
\end{align}
which characterise fluctuations of the operators $\sigma_{\pm}(t)$ around their steady-state values. The coherent contribution to the scattering thus gives rise to a $\delta$-function peak in the spectrum at the laser driving frequency, whereas the incoherent contribution may be calculated via the regression theorem as outlined above.

\begin{figure}[t!]
\begin{center}
\includegraphics[width=0.49\textwidth]{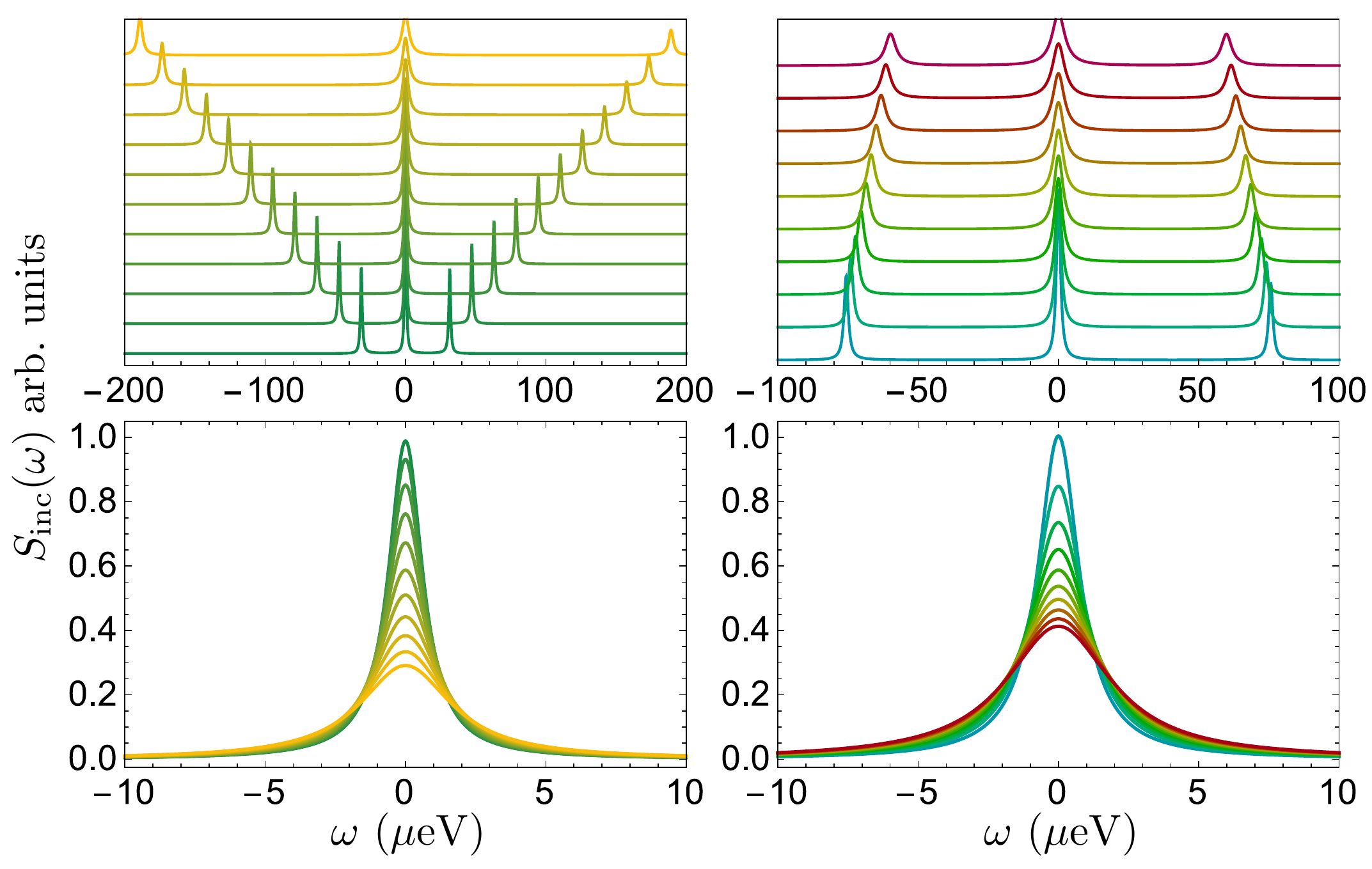}
\caption{Top left: QD emission spectra at $T=4$~K for varying driving strength, increasing from $\Omega=0.05$~ps$^{-1}$ to $\Omega=0.3$~ps$^{-1}$ in steps of $0.025$~ps$^{-1}$ (lower to upper spectra). Top right: QD emission spectra at $\Omega=0.12$~ps$^{-1}$ for varying temperature, increasing from $T=4$~K to $T=40$~K in steps of $4$~K (lower to upper spectra). The lower panels show the respective high-frequency sidebands plotted on top of each other to illustrate sideband broadening with driving strength (left) and temperature (right). 
Other parameters used: $\alpha=0.027~\mathrm{ps}^2$, $\w_c=2.2~\mathrm{ps}^{-1}$, $T_1=1/\gamma=700$~ps, 
and we drive the QD at the polaron-shifted resonance, $\delta=\int_0^{\infty}\mathrm{d}\w J_{\rm ph}(\w)/\w$.}
\label{spectra}
\end{center}
\end{figure}

The relative contributions of the coherent and incoherent components to the total scattered light may be found by integrating over frequency:
\begin{align}\label{contributions}
P_{\rm coh}&\propto\int_{-\infty}^{\infty}\mathrm{d}\omega S_{\rm coh}(\omega)={\rm lim}_{\tau\rightarrow\infty}[g^{(1)}(\tau)],\\
P_{\rm inc}&\propto\int_{-\infty}^{\infty}\mathrm{d}\omega S_{\rm inc}(\omega)=g^{(1)}(0)-{\rm lim}_{\tau\rightarrow\infty}[g^{(1)}(\tau)].
\end{align}
This allows us to define the fraction of coherent light as
\begin{align}\label{cohfraction}
{\mathcal F}_{\rm coh}=\frac{P_{\rm coh}}{P_{\rm coh}+P_{\rm inc}}=\frac{{\rm lim}_{\tau\rightarrow\infty}[g^{(1)}(\tau)]}{g^{(1)}(0)}=\frac{|\rho_{0X}(\infty)|^2}{\rho_{XX}(\infty)},
\end{align}
which is simply the ratio of the long and short time values of $g^{(1)}(\tau)$ (both of which are real).
In the absence of phonon interactions, for example for an atom in a cavity, the coherent fraction is a monotonically decreasing function with increasing Rabi frequency $\Omega$, and becomes strongly suppressed for $\Omega>\sqrt{2}\gamma$ 
as the TLS becomes saturated~\cite{stecknotes}. In the bottom right panel of Fig.~\ref{g1}, we see that for a QD in the presence of a phonon environment, the coherent fraction also decreases as the driving strength is increased from zero, with particularly strong suppression above $\sqrt{2}\gamma$. However, in stark contrast to the atomic case, once $\Omega\sim k_BT$ we surprisingly see a reemergence of coherent scattering, with almost half the light being coherently scattered at large driving strengths~\cite{McCutcheon2013}. Why should this be the case? Returning to Eq.~(\ref{cohfraction}), we see that the coherent fraction can be written as the ratio of the square of the absolute value of the QD coherence to the QD excitonic population, both calculated in the steady-state. Above saturation, the steady-state QD population is essentially unchanging, whereas the phonon bath attempts to thermalise the QD with respect to its internal Hamiltonian, given by $H_{\rm SV}$ of Eq.~(\ref{HSV}) in the variational representation. Furthermore, the phonon influence increases with driving strength, provided that $\Omega$ does not become much larger than the phonon cut-off frequency. In fact, neglecting emission and assuming a thermal state $\rho(\infty)\sim e^{-\beta \Omega_{\rm V}\sigma_x/2}/{\rm Tr [e^{-\beta \Omega_{\rm v}\sigma_x/2}}]$ under resonant driving, we obtain
\begin{align}
{\mathcal F}_{\rm coh}=\frac{|\rho_{0X}(\infty)|^2}{\rho_{XX}(\infty)}\sim\frac{1}{2}\tanh^2{\left(\frac{\Omega_{\rm v}}{k_BT}\right)},
\end{align}
which reaches a maximum value of $1/2$ for $\Omega_{\rm v}\gg k_BT$. This is also apparent from the dynamical evolution of $g^{(1)}(\tau)$ shown in Fig.~\ref{g1}, which relaxes to a finite steady-state value at both very weak and strong driving, but not in-between.

Turning now to the incoherent spectrum, in Fig.~\ref{spectra} we illustrate its behaviour with both varying driving strength and temperature. As can be seen, for sufficiently strong driving, a triple peak structure is evident in the incoherent spectrum, known as the Mollow triplet~\cite{mollow69}, with sidebands positioned at $\pm\Omega_{\rm v}$ around a central peak at the laser driving frequency. The Mollow triplet can be understood within the dressed state picture as arising from photon induced transitions between manifolds of the QD-laser dressed states, which are split by $\Omega_{\rm v}$. The sideband position increases linearly with $\Omega$ and each also broadens due to the associated enhancement of the phonon influence. This is shown explicitly in the lower left panel, where the high-frequency sidebands from the top left panel are plotted on top of each other to aid comparison. Interestingly, in the presence of phonon interactions, the incoherent spectrum also varies with temperature at constant driving strength, as shown in the right hand panels of Fig.~\ref{spectra}. Here we see a reduction in the sideband splitting as temperature increases, due to a suppression of the renormalised driving strength $\Omega_{\rm v}$, which has also been observed experimentally~\cite{Wei2014}. Again, for large enough temperature, significant sideband broadening can also be seen due to the thermal enhancement of phonon processes.

\section{Summary}
\label{conclusions}

We have reviewed, in some detail, master equation approaches to modelling the effects of exciton-phonon interactions in both the dynamics and emission spectra of optically driven QDs. We can summarise our conclusions as follows. A phenomenological pure-dephasing description can be useful for quick insight into QD population dynamics, but it must be used with care. A pure-dephasing approximation does not, for example, bring about the correct thermal equilibrium QD steady-state to which the phonon bath should naturally lead. This failure can be reflected, for example, in incorrect predictions for the QD coherence dynamics and emission spectra. More generally, we can say that processes induced by the phonon environment are not solely of a pure-dephasing form, unless the external driving strength goes to zero. Even in this case, a constant (Markovian) pure-dephasing rate is unable to capture the QD coherence dynamics brought about by phonon bath relaxation.

\begin{figure}[t!]
\begin{center}
\includegraphics[width=0.465\textwidth]{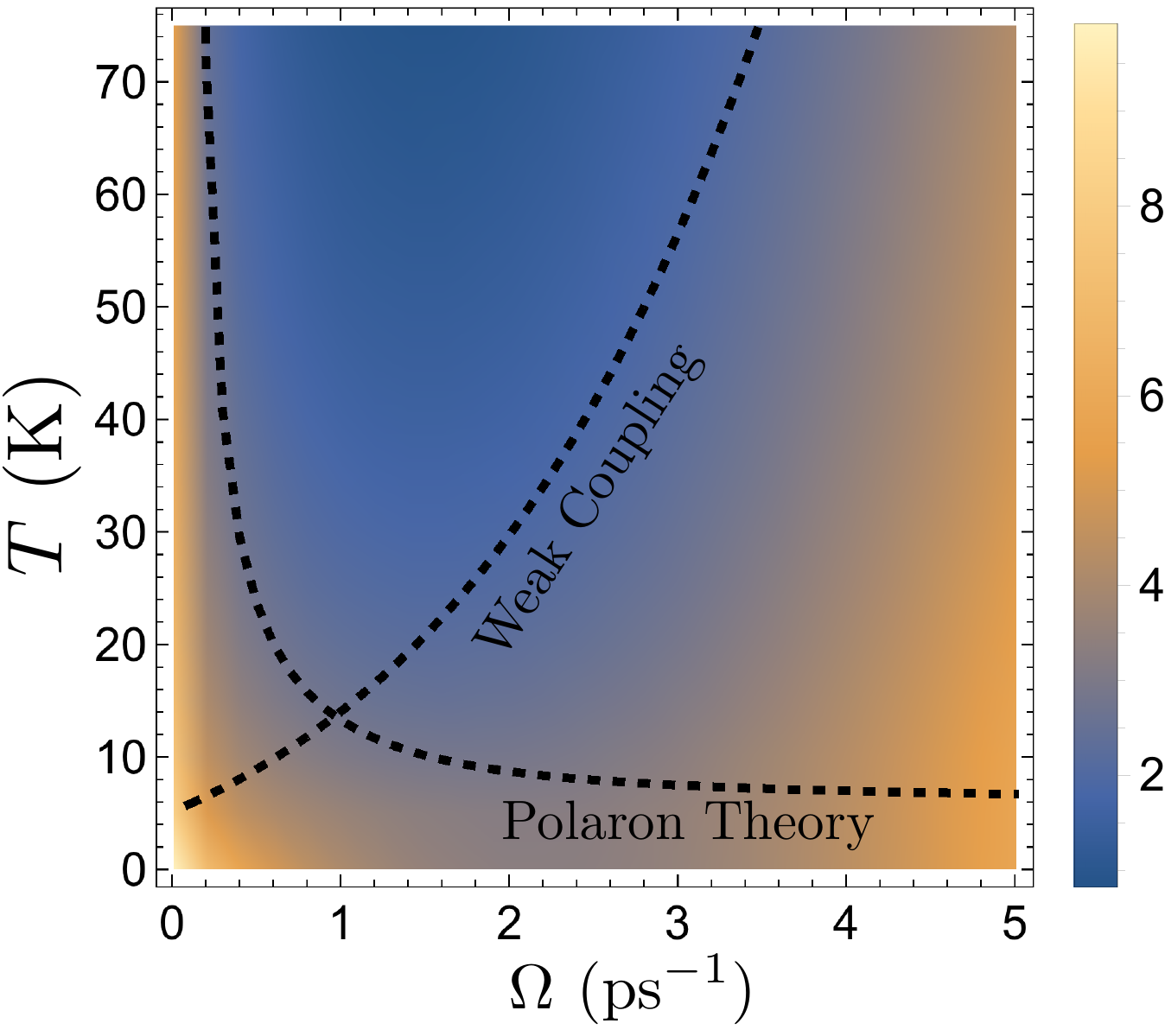}
\caption{Log (to base $e$) ratio of the QD population oscillation frequency to damping rate, calculated using the variational theory, as a function of temperature and Rabi frequency. Approximate regimes of validity for the weak-coupling and polaron approaches are also indicated. The lightest regions correspond to weakly damped coherent population oscillations in the time domain, while the darkest regions represent strongly damped dynamics. Other parameters used: $\alpha=0.027~\mathrm{ps}^2$, $\w_c=2.2~\mathrm{ps}^{-1}$, and we drive QD at the polaron-shifted resonance, $\delta=\int_0^{\infty}\mathrm{d}\w J_{\rm ph}(\w)/\w$.}
\label{PhaseDiagramTemp}
\end{center}
\end{figure}

Moving to a more microscopic approach, we may treat the exciton-phonon coupling term as a weak perturbation provided that single phonon processes dominate. However, at elevated temperatures and/or for strong exciton-phonon interactions such a weak-coupling treatment will break down, and it may even predict non-physical dynamics for quite reasonable experimental parameters. More specifically, perturbation in the exciton-phonon interaction cannot properly describe multiphonon QD Rabi frequency renormalisation effects due to the presence of the phonon bath, nor the resulting changes to the QD damping rates~\cite{mccutcheon10_2}.

To go beyond weak-coupling, it is possible to formulate a master equation within the polaron representation, which accounts for QD-state-dependent phonon displacement (i.e~polaron formation) at the Hamiltonian level. The resulting theory properly captures QD Rabi frequency renormalisation, and multiphonon processes more generally, provided that the phonon modes can adiabatically follow the QD state. This means that it is applicable at weak to strong couplings and low to high temperatures. However, if the external optical driving is so strong that the phonon modes become sluggish (i.e.~low in frequency compared to the Rabi scale), they are unable to track the state of the QD. The polaron basis is then no longer appropriate and the theory fails. In particular, it is generally unable to reproduce the expected decoupling effect between the exciton and phonons at very strong driving, which is a feature of both weak-coupling~\cite{nazir08} and numerically converged path-integral calculations~\cite{vagov2007}.

\begin{figure}[t!]
\begin{center}
\includegraphics[width=0.495\textwidth]{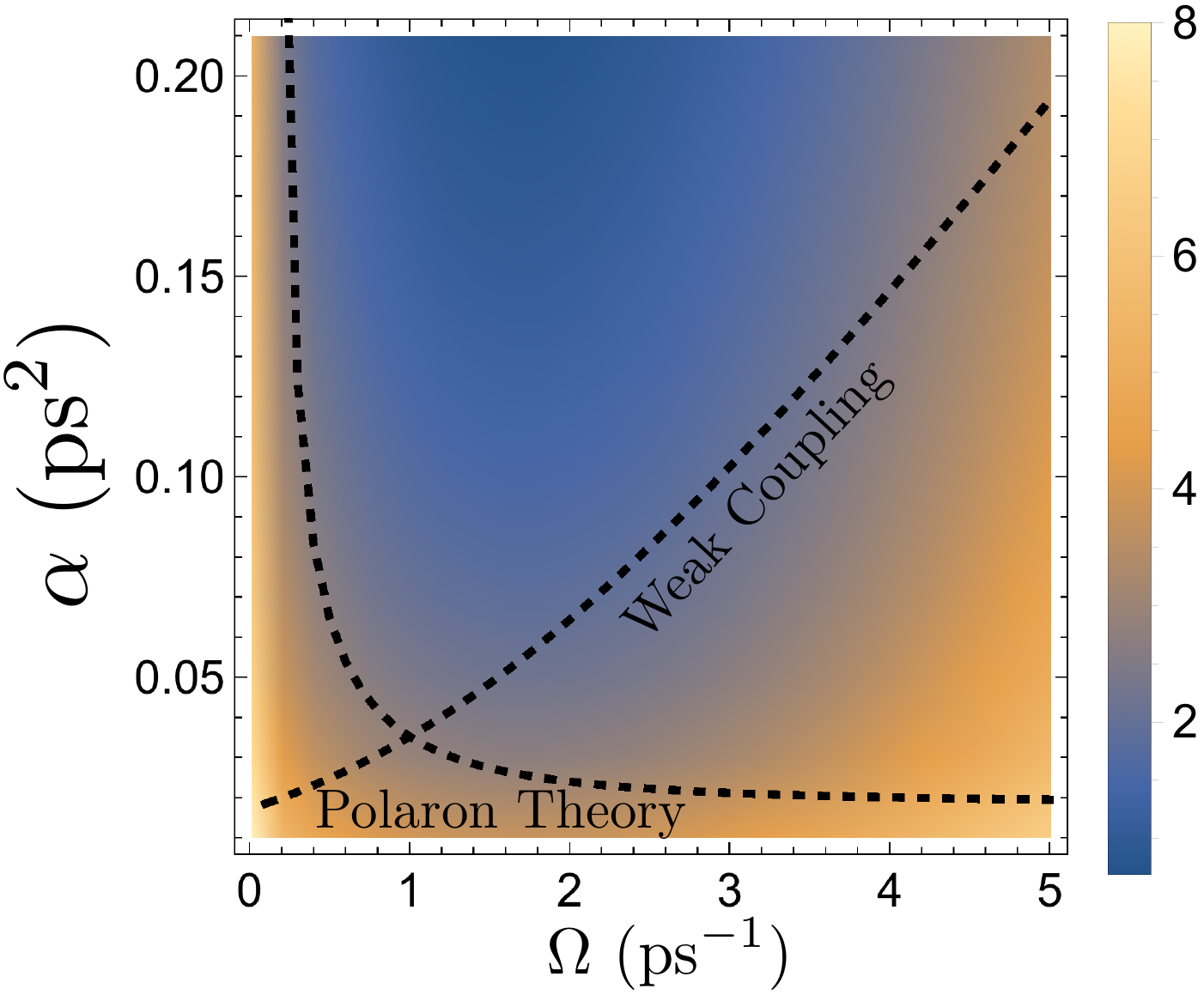}
\caption{Log (to base $e$) ratio of the QD population oscillation frequency to damping rate, calculated using the variational theory, as a function of exciton-phonon coupling strength and Rabi frequency. Approximate regimes of validity for the weak-coupling and polaron approaches are also indicated. The lightest regions correspond to weakly damped coherent population oscillations in the time domain, while the darkest regions represent strongly damped dynamics. Other parameters used: $T=10~\mathrm{K}$, $\w_c=2.2~\mathrm{ps}^{-1}$, and we drive QD at the polaron-shifted resonance, $\delta=\int_0^{\infty}\mathrm{d}\w J_{\rm ph}(\w)/\w$.}
\label{PhaseDiagramCoupling}
\end{center}
\end{figure}

The most versatile approach explored is also based on applying a transformation of polaron form, but now with phonon mode dependent displacements determined by a variational procedure. This theory is flexible enough to encompasses both the weak-coupling and polaron methods in the appropriate limits, and can also accurately interpolate between them~\cite{mccutcheon11_2}. The variational optimisation naturally chooses a representation that, as far as possible given the specific form of transformation, minimises the perturbative terms within the Hamiltonian. The resulting master equation thus remains valid over a much larger regime of parameter space as compared to less sophisticated techniques.

These considerations are captured, broadly speaking, in Figs.~\ref{PhaseDiagramTemp} and~\ref{PhaseDiagramCoupling}. Here, by fitting the variational theory QD population dynamics to an exponentially damped oscillation we have extracted a Q-factor, which is the ratio of the oscillation frequency to the damping rate. We plot this on a log (to base $e$) scale as a function of temperature and Rabi frequency (at fixed exciton-phonon coupling) in Fig.~\ref{PhaseDiagramTemp} and as a function of exciton-phonon coupling strength and Rabi frequency (at fixed temperature) in Fig.~\ref{PhaseDiagramCoupling}. We have also indicated a rough region of validity for the full polaron theory by comparing the polaron ($\Omega_{\rm p}$) and variational ($\Omega_{\rm v}$) renormalisations. The dashed curve indicates a $5$~\% difference in renormalisation, and below this we expect the full polaron theory to be accurate. Similarly, we use a $5$~\% deviation of $\Omega_{\rm v}$ from the unrenormalised $\Omega$ to define an approximate region of validity for the weak-coupling treatment, below which multiphonon effects are of minor importance. We note that for typical driving frequencies in continuous-wave experiments ($\Omega\ll\omega_c$), the polaron and variational approaches open up the possibility to explore a much larger range of exciton-phonon coupling strengths and phonon bath temperatures than a weak-coupling treatment.

Finally, we mention that the techniques we have reviewed are not restricted to the study of QD exciton-phonon interactions, but may be applied much more widely to quantum systems coupled to bosonic environments~\cite{wurger98,aslangul85,aslangul86,dekker87,chin2011generalized,nazir12}. In particular, the polaron and variational approaches have recently gained substantial interest in relation to excitonic energy transfer in molecular dimers~\cite{nazir09,jang08,jang09,mccutcheon11,mccutcheon11_3,zimanyi12} and larger light-harvesting complexes~\cite{kolli11,jang11,kolli12,pollock13}. Here, the exciton-vibrational coupling can naturally be quite strong and the spectral density may take on a more complicated, structured form. In fact, in such situations, system-environment correlations and non-Markovian dynamics can become especially important~\cite{roden09,prior10,ishizaki10,nalbach11,cerrillo14,PhysRevA.90.032114}, with the question of how 
to extend master equation approaches to incorporate such effects a key driver for a very active and important area of research.   



\appendix

\section{Calculation of correlation functions using coherent states}

\subsection{Coherent states}\label{coherentstates}

Let us begin by defining and reviewing some important properties of coherent states, for which  
we follow the seminal work of Glauber~\cite{glauber63}. 
Coherent states constitute an alternative to the Fock basis used to describe states of 
harmonic oscillators. Let us consider a single harmonic oscillator mode described by 
creation and annihilation operators $b^{\dagger}$ and $b$ satisfying $[b,b^{\dagger}]=1$, 
with 
frequency $\w$. A Fock state $\ket{n}$ 
is defined as an eigenstate of the number operator $b^{\dagger}b$ such that 
\beq
b^{\dagger}b \ket{n} = n\ket{n}.
\eeq
If the oscillator is not coupled to any other system its Hamiltonian is $H = (\w+1/2) b^{\dagger} b$ 
and therefore $\ket{n}$ is a state with definite energy $(n+1/2)\w$. From these 
basic definitions it can be shown that the Fock states also satisfy
\beq
b \ket{n}= \sqrt{n} \ket{n-1},
\qquad b^{\dagger} \ket{n} =\sqrt{n+1} \ket{n+1},
\eeq
and that $\braket{n}{m}=\delta_{nm}$.

Coherent states are defined as eigenstates of the annihilation operator. We label a coherent 
state 
as $\ket{\alpha}$, which satisfies 
\beq
b \ket{\alpha}=\alpha\ket{\alpha},
\eeq
where $\alpha$ is some complex number referred to as the amplitude of the coherent state. 
From this basic definition it can be 
shown that a coherent state has a Fock state representation 
\beq
\ket{\alpha}=\e^{-\frac{1}{2}|\alpha|^2}\sum_n \frac{\alpha^n}{\sqrt{n!}}\ket{n},
\eeq
from which it follows that $\braket{0}{\alpha}=\e^{-\frac{1}{2}|\alpha|^2}$ with $\ket{0}$ the vacuum satisfying $b \ket{0} = 0$. 
To generate the coherent states, we consider the action of the displacement operator 
\beq
D(\alpha)=\e^{\alpha b^{\dagger}-\alpha^* b}=\e^{-\frac{1}{2}|\alpha|^2}\e^{\alpha b^{\dagger}}\e^{-\alpha^* b},
\eeq
where the second equality follows from the identity $\e^{A+B}=\e^{-\frac{1}{2}[A,B]}\e^A \e^B$, valid 
when $[A,B]$ is proportional to the identify. From this definition 
it can be seen that 
\beq
D(\alpha)\ket{0}=\ket{\alpha}.
\eeq
%
The displacement operator is named as 
such since it transforms $b^{\dagger}$ and $b$ according to 
\begin{align}
D(\alpha) b^{\dagger} D^{\dagger}(\alpha)&=b^{\dagger}-\alpha^*,\label{bdagDisplaced}\\
D(\alpha) b D^{\dagger}(\alpha)&=b-\alpha.\label{bDisplaced}
\end{align}
It is also useful to note that displacement operators can be combined using the relation 
\beq
D(\alpha_2)D(\alpha_1)=D(\alpha_2+\alpha_1)\e^{\frac{1}{2}(\alpha_2 \alpha_1^*-\alpha_2^* \alpha_1)}.
\label{DisplacementCombination}
\eeq
%

In order to appreciate the utility of the coherent state representation we now consider 
how it can be used to calculate expectation values. 
The key ingredient is to notice that the identity can be expressed in terms of 
coherent states as 
\beq
\frac{1}{\pi}\int \mathrm{d}^2 \alpha |\alpha\rangle\langle\alpha|= \openone,
\eeq
%
%
where the integral takes place over the whole complex plane, i.e.~$\int \mathrm{d}^2 \alpha = \int_{-\infty}^{\infty}\mathrm{d}\mathrm{Re}[\alpha]\int_{-\infty}^{\infty}\mathrm{d}\mathrm{Im}[\alpha]$. 
With this identification it can be shown that the trace of an operator can be written 
\beq
\mathrm{Tr}[ A ] = \frac{1}{\pi}\int \mathrm{d}^2 \alpha\langle\alpha|A\ket{\alpha}.
\eeq
We can write a density operator as 
\beq
\rho = \int \mathrm{d}^2 \alpha P(\alpha) |\alpha\rangle\langle\alpha|,
\eeq
where $P(\alpha)$ satisfies $\int \mathrm{d}^2 \alpha P(\alpha)=1$ and describes the state. It follows that the expectation 
value of an operator $A$ with respect to the density operator described by $P(\alpha)$ is given by 
\beq
\av{A}=\mathrm{Tr}[A \rho]=\int\mathrm{d}^2\alpha P(\alpha)\langle\alpha|A\ket{\alpha}.
\eeq
We are typically interested in thermal state density operators, for which it can be shown 
that $P(\alpha)=(1/\pi N)\exp(-|\alpha|^2/N)$ where $N = (\e^{\beta \w}-1)^{-1}$ is the average number of 
excitations in an oscillator of frequency $\w$ at inverse temperature $\beta=1/k_B T$.

\subsection{Correlation functions}\label{CorrelationFunctions}

We are now in a position to calculate the bath correlation functions used in the derivation of our master equations 
in the main text. We begin with the weak-coupling correlation function encountered in Section~\ref{WeakCoupling}. 
Considering for now the single mode case, we are interested in the expectation value 
%
\beq
\av{\tilde{B}_z(\tau)\tilde{B}_z(0)}=\!
\frac{1}{\pi N}\!\int\!\!\mathrm{d}^2 \alpha\, \e^{-|\alpha|^2/N}\langle\alpha|\tilde{B}(\tau)\tilde{B}(0)\ket{\alpha},
\eeq
where 
$\tilde{B}_z(\tau)=g(b^{\dagger}\e^{i \w \tau}+b\e^{-i \w \tau})$. By writing 
$\ket{\alpha}=D(\alpha)\ket{0}$ and inserting $\openone=D(\alpha)D(-\alpha)$ in-between the two factors 
of $B_z$, we may use Eqs.~({\ref{bdagDisplaced}}) and ({\ref{bDisplaced}}) 
and perform the necessary complex integrals to find 
\beq
\av{\tilde{B}_z(\tau)\tilde{B}_z(0)}=g^2\big(\cos\w\tau\coth(\beta\w/2)-i\sin\w\tau\big).
\label{CWeakSingleMode}
\eeq
As the modes are independent, generalising to the multimode case results in a sum of terms of the form in 
Eq.~({\ref{CWeakSingleMode}}), and converting the sum to an integral with use of the 
spectral density results in Eq.~({\ref{CWeak}}) in the main text.

In the polaron theory we have correlation functions of displacement operators. We first consider 
the expectation value of a single displacement operator, $\av{B_{\pm}}$, which has the form 
\beq
\av{D(h)}
=\frac{1}{\pi N}\int\mathrm{d}^2 \alpha \,\e^{-|\alpha|^2/N}\langle\alpha|D(h)\ket{\alpha}, 
\eeq
where we take $h = \pm (g/\w)$. 
Once again we use $D(\alpha)\ket{0}=\ket{\alpha}$, and with the aid of Eq.~({\ref{DisplacementCombination}}) 
we find after integration
\beq
\av{D(h)}=\exp\Big[-\frac{1}{2}|h|^2 \coth(\beta\w/2)\Big].
\label{avD}
\eeq
Generalising to the multimode case we obtain the polaron theory renormalisation factor given in Eq.~({\ref{Bav}}). 

Now, in order to calculate correlation functions we must consider the form  
$\av{D(h)D(h')}$ for some complex numbers $h$ and 
$h'$. For this we can simply use Eq.~({\ref{DisplacementCombination}}) to write 
$\av{D(h)D(h')}=\exp[\frac{1}{2}(h h'^*-h^*h')]\av{D(h+h')}$. 
Together with Eq.~({\ref{avD}}) it can be seen that $\av{D(\pm h)D(\mp h')}=\av{D(\mp h)D(\pm h')}$ 
while $\av{D(\pm h)D(\pm h')}=\av{D(\mp h)D(\mp h')}$. 
Now, letting $h = (g/\w)\e^{i \w \tau}$ and $h' = (g/\w)$ 
the time dependent correlation functions can readily be found to be 
\begin{align}
\av{\tilde{B}_{\pm}(\tau) \tilde{B}_{\mp}(0)}&=\av{B}\e^{\phi(\tau)},\label{Bpmappendix}\\
\av{\tilde{B}_{\pm}(\tau) \tilde{B}_{\pm}(0)}&=\av{B}\e^{-\phi(\tau)},\label{Bppappendix}
\end{align}
where in the single mode case the phonon propagator is 
\beq
\phi(\tau)=\left(\frac{g}{\w}\right)^2\!\!\big(\cos\w \tau \coth(\beta\w/2)-i \sin\w \tau\big).
\eeq
Recalling the definitions of the multimode bath operators in the polaron theory from Eq.~({\ref{BxBy}}) 
we arrive at the correlation functions given in Eqs.~({\ref{Cxx}}) and ({\ref{Cyy}}).

In the variational theory the interaction Hamiltonian has two types of bath operator; 
one being essentially $B_z$ encountered in the weak-coupling theory, but 
with the coupling constant $g$ replaced with $g-f$, and 
the other being the same as in polaron theory but with $g$ replaced by $f$ 
[see Eqs.~({\ref{BxByVar}}) and ({\ref{BzVar}})]. As such, the master equation 
involves three types of correlation function, one involving two weak-coupling like operators, 
one involving polaron theory-like operators, and a third type unique to the variational 
theory involving cross terms. To evaluate the cross terms we consider 
\begin{align}
\av{\tilde{\mathcal{B}}_{\pm}(\tau) \tilde{\mathcal{B}}_z(0)}
&=\frac{1}{\pi N}\nonumber\\
&\times\int\!\!\mathrm{d}^2 \alpha\,\e^{-|\alpha|^2/N} {} \langle \alpha | D(h) (g - f)(b^{\dagger}+b)|\alpha \rangle,
\end{align}
where we have $h = \pm (f/\w)\e^{i\w \tau}$ for $\tilde{\mathcal{B}}_{\pm}(\tau)$, respectively. 
Using the usual trick of writing $|\alpha\rangle = D (\alpha)\ket{0}$, permuting 
$D (\alpha)$ through $\smash{(b^{\dagger}+b)}$ with the use of 
Eqs.~({\ref{bdagDisplaced}}) and ({\ref{bDisplaced}}), and combining the three displacement operators 
we arrive at 
\begin{align}
\av{\tilde{\mathcal{B}}_{\pm}(\tau) \tilde{\mathcal{B}}_z(0)}
=(g - f)
\av{D(h)}(N h - (N+1)h^*).
\end{align}
Recalling Eq.~({\ref{BxByVar}}) relating the variational operators 
$\mathcal{B}_x$ and $\mathcal{B}_y$ to $\mathcal{B}_{\pm}$ in the variational theory, 
we find 
\begin{align}\label{Cyzsinglemode}
\av{\tilde{\mathcal{B}}_{y}(\tau) \tilde{\mathcal{B}}_z(0)}
=&-\av{\mathcal{B}}(g - f)\Big(\frac{f}{\w}\Big)\nonumber\\
&\times
\big(i \cos\w \tau + \sin\w\tau\coth(\beta\w/2)\big),
\end{align}
while $\av{\tilde{\mathcal{B}}_{z}(\tau) \tilde{\mathcal{B}}_y(0)}=-\av{\tilde{\mathcal{B}}_{y}(\tau) \tilde{\mathcal{B}}_z(0)}$, and 
$\av{\tilde{\mathcal{B}}_{z}(\tau) \tilde{\mathcal{B}}_x(0)}=\av{\tilde{\mathcal{B}}_{x}(\tau) \tilde{\mathcal{B}}_z(0)}=0$.
The generalisation of Eq.~(\ref{Cyzsinglemode}) to the multimode case is given in Eq.~(\ref{cyzvar}).


%

\end{document}